\documentclass[aps,pra,twocolumn,groupedaddress,nofootinbib]{revtex4-1}
\usepackage[english]{babel}
\usepackage{amsmath}
\usepackage{amsfonts}
\usepackage{amssymb}
\usepackage{amsthm}
\usepackage{graphicx}

\newcommand{\h}{H^{(+)}}
\newcommand{\hp}{H^{(+)\prime}}
\newcommand{\hh}{H^{(-)}}

\begin{document}

\title{Tutorial\\ Lorenz-Mie theory for 2D scattering and resonance calculations}
\author{Denis Gagnon}
\author{Louis J. Dub\'e}
\date{\today}
\address{D\'epartement de physique, de g\'enie physique et d'optique \\Facult\'e des Sciences et de G\'enie, Universit\'e Laval, Qu\'ebec, G1V 0A6, Canada}

\email{Corresponding author: ljd@phy.ulaval.ca}

\begin{abstract}

This PhD tutorial is concerned with a description of the two-dimensional generalized Lorenz-Mie theory (2D-GLMT), a well-established numerical method used to compute the interaction of light with arrays of cylindrical scatterers.
This theory is based on the method of separation of variables and the application of an addition theorem for cylindrical functions.
The purpose of this tutorial is to assemble the practical tools necessary to implement the 2D-GLMT method for the computation of scattering by passive scatterers or of resonances in optically active media.
The first part contains a derivation of the vector and scalar Helmholtz equations for 2D geometries, starting from Maxwell's equations.
Optically active media are included in 2D-GLMT using a recent stationary formulation of the Maxwell-Bloch equations called \emph{steady-state ab initio laser theory} (SALT), which introduces new classes of solutions useful for resonance computations.
Following these preliminaries, a detailed description of 2D-GLMT is presented.
The emphasis is placed on the derivation of beam-shape coefficients for scattering computations, as well as the computation of resonant modes using a combination of 2D-GLMT and SALT.
The final section contains several numerical examples illustrating the full potential of 2D-GLMT for scattering and resonance computations.
These examples, drawn from the literature, include the design of integrated polarization filters and the computation of optical modes of photonic crystal cavities and random lasers. 

\end{abstract}

%\pacs{}

\maketitle

\tableofcontents

\section{Introduction}

Controlling the flow of light on the micro- and nano-scale level is one of the main technological breakthroughs enabled by the use of complex photonic media \cite{Noginov2009}.
These complex media can be viewed as composite optical materials whose characteristic length scale is of the order of the operating wavelength.
Some prime examples of complex photonic media are photonic crystals \cite{Aryal2009, DeLaRue2012}, metamaterials \cite{Cui2014} as well as random lasers \cite{Wiersma2008, Cao2009}.
Besides the fact that the feasibility of all these examples has been demonstrated experimentally, a large body of work concerning the \emph{numerical modelling} of complex photonic media has accumulated over the years.
For several reasons, the development of these numerical tools is a challenging research topic in itself.
Firstly, most of the tools typically taught in graduate courses in optics, such as the paraxial approximation and $ABCD$ matrix approaches, are often incompatible with wavelength-scale structures.
Additionally, novel photonic media often involve a plethora of physical phenomena which can only be addressed numerically.
A non-exhaustive list includes bandgaps \cite{Aryal2009, DeLaRue2012}, resonances \cite{Lukyanchuk2010}, slow light \cite{Baba2008}, negative refraction \cite{McCall2009}, topological insulation \cite{Khanikaev2013} and Anderson localization of light \cite{Riboli2014}.
Accordingly, numerical modelling tools aimed at complex photonic media must often compose with the full wave nature of the electromagnetic (EM) field (described by Maxwell's equations), potentially active or non-linear media, as well as non-trivial boundary conditions of sub-wavelength geometries.

Numerical schemes for the solution of Maxwell's equations that satisfy the above requirements can be classified in two broad categories.
The first category of methods is based on spatial and temporal discretization of the underlying differential equations \cite{Sadiku2000, Taflove2005}.
Well known examples include finite element methods (FEM) as well as finite-difference-time-domain approaches (FDTD).
Although these fully numerical schemes are very general, meshing is most often associated with heavy memory requirements, especially in the case where the spatial extent of the complex geometry considered is several times larger than the operating wavelength.
In contrast, the second category of methods is based on the expansion of the EM field over a set of basis functions suited to specific geo\-metries.
In other words, these methods exploit an underlying geometrical symmetry of the problem.
One obvious advantage of using a natural basis of the problem is that it alleviates the numerical difficulties associated with the first category, since it does not imply a spatial discretization of the problem space.
The main topic of this tutorial is a method of the second category, which we refer to as \emph{two-dimensional generalized Lorenz-Mie theory} (2D-GLMT) \cite{Gouesbet2013}.
This method is best suited for the modelling of effectively 2D complex photonic media based on arrangements of disconnected and parallel cylindrical scatterers.
These structures include, for instance, photonic devices based on coupled scatterers in the form of cylindrical holes in a dielectric slab waveguide, as well as devices based on coupled pillars \cite{Xu2010} or coupled circular cavities \cite{Nakagawa2005, Ishii2006}.
These effectively 2D devices are of great practical interest since they can easily be integrated on photonic circuits.
This allows the fabrication of innovative optical elements onto existing technological platforms, for instance silicon-on-insulator wafers \cite{DeLaRue2012, Rickman2014} or plastic substrates \cite{Li2014}.

The purpose of this tutorial is to give a pedagogical, hands-on introduction to the theory behind 2D-GLMT, the main numerical aspects of the method as well as its wide range of applications.
It is intended for graduate students or researchers wishing to construct their own implementation of the method and apply it to their research projects.
The applications targeted in this tutorial are varied, but the emphasis is placed on two classes of optical geometries. 
The first is \emph{photonic-crystal inspired devices}, designed with the conventional wisdom of 2D photonic crystals (PhCs) in mind, but which are not necessarily infinite nor periodic.
This research field has been active since the seminal 1987 publication by Yablonovitch, in which the author observed that multiple scattering and interference processes in a PhC could slow down the group velocity of light to the point of completely inhibiting propagation \cite{Yablonovitch1987}.
Frequency intervals in which propagation is inhibited are called \emph{bandgaps}, following the analogy with electronic bandgaps in semi-conductor crystals.
Nowadays, PhC-inspired optical devices can readily be engineered using micro-fabrication methods such as UV \cite{Xu2010}, holographic \cite{Skorobogatiy2009} or electron beam \cite{DeLaRue2012, Riboli2014} lithography.
This relative ease of fabrication has enabled the realization of various integrated optical elements exploiting bandgaps in PhCs, including Mach-Zehnder interferometers \cite{Nordin2002, Pottier2006}, waveguides \cite{Jin2005, Xing2005}, mirrors \cite{Pottier2006}, lenses \cite{Sanchis2004, Marques-Hueso2013}, polarization beam-splitters \cite{Pottier2006}, beam shapers \cite{Gagnon2012} and all-optical switches \cite{Heuck2013}.

The second class of applications is called \emph{complex lasing media}, and include PhC lasers \cite{Painter1999}, photonic atoms and molecules \cite{Boriskina2006, Boriskina2010a} as well as random lasers \cite{Andreasen2011, Cao2009, Tuereci2008, Tuereci2009}.
Although PhC lasers could fall in the first class of applications, we make the distinction since there are subtle differences between using 2D-GLMT for modelling passive devices and its use to compute lasing modes.
Clearly these two classes do not exhaust the list of possible applications of 2D-GLMT. 
Indeed, the method could be also extended to any optical geometry constituted of coupled homogeneous cylinders, for instance micro-structured hollow-core fibres \cite{White2002, Kuhlmey2002, Noble2011}.

The tutorial is organized as follows. In section \ref{sec:em}, we review the basic concepts of electromagnetism relevant to 2D-GLMT.
We begin by deriving the vector and scalar Helmholtz equations for coupled cylindrical geometries, starting from Maxwell's equations. 
After a derivation of the Helmholtz equations for passive optical media, we present a description of complex lasing media based on a stationary formulation of the Maxwell-Bloch equations called \emph{steady-state ab initio laser theory} (SALT).
This recently formulated approach, as well as the new kind of optical eigenmodes it introduces, are described in section \ref{sec:salt}.

Following these derivations of the basic electromagnetic equations for passive and active cylinders, we move on to a detailed description 2D-GLMT in section \ref{sec:glmt}.
This includes the derivation of expansion coefficients for two incident excitations, namely a plane wave (section \ref{sec:plane}) and a collimated quasi-Gaussian beam (section \ref{sec:csb}).
The types of computations that can be achieved using 2D-GLMT fall in two broad categories, which we refer to as ``scattering'' and ``eigenmodes'', respectively.
Since the objective is to describe as precisely as possible how to construct a numerical implementation of the method, examples of both types of computations are discussed in section \ref{sec:examples}.
Specifically, PhC-inspired devices dedicated to beam shaping and polarization filtering are presented in section \ref{sec:filter} as a representative example of scattering computations.
Photonic crystal cavities (section \ref{sec:phc}) and random lasers (section \ref{sec:random}) serve thereafter as examples of eigenmode computations.

\section{Electromagnetic theory}
\label{sec:em}
Our starting point is Maxwell's equations which, for a non-magnetic medium (unit magnetic permeability, or $\mu = 1$), take the form
\begin{subequations}
\begin{gather}
\nabla \cdot \mathbf{D} = \rho, \label{eq:deplacement} \\
\nabla \cdot \mathbf{H} = 0, \\
\nabla \times \mathbf{E} = -\frac{1}{c}\frac{\partial \mathbf{H} }{\partial t}, \label{eq:ne} \\
\nabla \times \mathbf{H} = \frac{1}{c}\frac{\partial  \mathbf{D} }{\partial t} + 
\frac{ \mathbf{J} }{c}. \label{eq:nh} 
\end{gather}
\end{subequations}
Without loss of generality, the system of Heaviside-Lorentz electromagnetic units is used throughout this work except where indicated. Details of the conversion between Heaviside-Lorentz and other systems, i.e. SI or Gaussian units, can be found in \cite{Desloge1994}.

We are mainly concerned with obtaining rigorous solutions of Maxwell's equations in regions of space that contain no free charges and no free currents, that is \cite{Boyd2003}
\begin{subequations}
\begin{gather}
\rho = 0, \\
\mathbf{J} = 0.
\end{gather}
\end{subequations}
However, since we are also interested in solutions of Maxwell's equations in \emph{active} media, we suppose the following relation between the displacement field $\mathbf{D}$ and the electric field $\mathbf{E}$ \cite{Boyd2003,Desloge1994}
\begin{equation}\label{eq:d}
\mathbf{D} = \mathbf{E} + \mathbf{P}_\mathrm{T},
\end{equation}
where $\mathbf{P}_\mathrm{T}$ is the total polarization field.
It contains both a contribution from the linear response of the medium (basically the optical density, or refractive index) and the non-linear response.
Taking the curl of \eqref{eq:ne} and \eqref{eq:nh}, together with \eqref{eq:d}, one obtains the following wave equations
\begin{subequations}
\begin{gather}
\nabla \times \nabla \times \mathbf{E} + \frac{1}{c^2}\frac{\partial^2 \mathbf{E} }{\partial t^2} 
+ \frac{1}{c^2}\frac{\partial^2 \mathbf{P}_\mathrm{T} }{\partial t^2} 
= 0, \label{eq:nne} \\
\nabla \times \nabla \times \mathbf{H} + \frac{1}{c^2}\frac{\partial^2  \mathbf{H} }{\partial t^2} - \frac{1}{c}\frac{\partial }{\partial t} \big[ \nabla \times \mathbf{P}_\mathrm{T} \big] 
= 0. \label{eq:nnh}
\end{gather}
\end{subequations}

\subsection{Dielectric media}

The two fundamental wave equations for $\mathbf{E}$ and $\mathbf{H}$ can be further reduced if the response of the optical media is assumed to be linear.
In this case, the polarization field $\mathbf{P}_\mathrm{T}$ is a linear function of the electric field, more specifically
\begin{equation}
\mathbf{P}_\mathrm{T} = \chi_e \mathbf{E}, \label{eq:linear}
\end{equation}
where $\chi_e$ is the electric susceptibility of the medium, i.e. $\chi_e = 0$ in free space. The substitution of the Ansatz \eqref{eq:linear} in \eqref{eq:nne} and \eqref{eq:nnh} yields, after a few vector identities,
\begin{subequations}
\begin{gather}
\nabla (\nabla \cdot \mathbf{E} ) - \nabla^2 \mathbf{E} 
+ \frac{\epsilon}{c^2}\frac{\partial^2 \mathbf{E} }{\partial t^2} = 0, \label{eq:nnelinear} \\
\nabla (\nabla \cdot \mathbf{H} ) - \nabla^2 \mathbf{H}
+ \frac{\epsilon}{c^2}\frac{\partial^2  \mathbf{H} }{\partial t^2} - \frac{1}{c}\frac{\partial }{\partial t} \big[ \mathbf{E} \times \nabla \chi_e \big]
= 0, \label{eq:nnhlinear}
\end{gather}
\end{subequations}
where $\epsilon \equiv 1 + \chi_e$ denotes the \emph{relative dielectric permittivity}.
When $\epsilon(\mathbf{r})$ is a continuous function of space, these equations can not be further simplified. 
However, in the case of arrays of individually homogeneous cylinders, $\epsilon(\mathbf{r})$ is a piecewise constant function. 
In other words, we will be concerned with regions of constant relative permittivity separated by well-defined interfaces, where the value of  $\epsilon(\mathbf{r})$ is discontinuous. 
It is then possible to solve the equations under the assumption that $\nabla \epsilon(\mathbf{r}) = 0$ and $\nabla \chi_e(\mathbf{r}) = 0$, and connect the solutions at the cylinders' interfaces using appropriate electromagnetic boundary conditions. 
Under these conditions, \eqref{eq:deplacement} reads
\begin{equation}
\nabla \cdot \mathbf{D} = \nabla \cdot [ \epsilon \mathbf{E} ] 
= \epsilon \nabla \cdot \mathbf{E} + \mathbf{E} \cdot \nabla \epsilon = 0
\end{equation}
and since $\nabla \epsilon(\mathbf{r}) = 0$, one has immediately that $\nabla \cdot \mathbf{E}=0$. 
This simplifies \eqref{eq:nnelinear} and \eqref{eq:nnhlinear} to the vector Helmholtz equations
\begin{subequations}
\begin{gather}
\nabla^2 \mathbf{E} 
- \frac{\epsilon (\mathbf{r}) }{c^2}\frac{\partial^2 \mathbf{E} }{\partial t^2} = 0, \\
\nabla^2 \mathbf{H}
- \frac{\epsilon (\mathbf{r}) }{c^2}\frac{\partial^2  \mathbf{H} }{\partial t} = 0.
\end{gather}
\end{subequations}
These equations are more conveniently solved in the frequency domain, using the following substitution for temporal derivatives
\begin{equation}
\frac{\partial^2}{\partial t^2} \leftarrow - \omega^2.
\end{equation}
This is of course the assumption of a harmonic time-dependence of the electromagnetic fields, i.e. $\mathbf{E} = \mathbf{E}(\mathbf{r})e^{-i\omega t}$ and $\mathbf{H} = \mathbf{H}(\mathbf{r})e^{-i\omega t}$.
This last step yields
\begin{subequations}
\begin{gather}
\nabla^2 \mathbf{E} 
+ \epsilon(\mathbf{r}) k^2 \mathbf{E} = 0, \label{eq:vectore} \\
\nabla^2 \mathbf{H}
+ \epsilon(\mathbf{r}) k^2 \mathbf{H} = 0, \label{eq:vectorh}
\end{gather}
\end{subequations}
where the wavenumber $k \equiv \omega / c$.

The vector Helmholtz equation is a general description of wave propagation in fully three-di\-men\-sional, linear media of piecewise constant refractive index ($\nabla \epsilon(\mathbf{r}) = 0$).
This covers a wide range of optical geometries: optical fibres and waveguides \cite{Okamoto2006}, dielectric cavities such as spheres and tori \cite{Vahala2003} and three-dimensional PhCs \cite{Aryal2009, Maigyte2013, Skorobogatiy2009} to name a few.

Since the scope of this tutorial is cylindrical geometries, an additional symmetry can be used to further reduce the vector Helmholtz equation to a scalar form. 
Effectively two dimensional geometries suggest two privileged polarization directions. 
Formally, it is possible to uncouple the electromagnetic field in two orthogonal polarization components if variations of the relative permittivity $\epsilon(\mathbf{r})$ are restricted to a plane, say the $(x,y)$ plane, meaning 
\begin{equation}
  \epsilon(\mathbf{r}) = \epsilon(x,y).
 \end{equation}
This 2D hypothesis implies that the electromagnetic field can be uncoupled in a transverse-electric (TE) component ($\mathbf{E}$ field component parallel to the $x,y$ plane, $\mathbf{H}$ field component normal to the $x,y$ plane) and a transverse magnetic (TM) component ($\mathbf{H}$ field component parallel to the $x,y$ plane, $\mathbf{E}$ field component normal to the $x,y$ plane). 
Under this two-dimensional restriction, \eqref{eq:vectore} and \eqref{eq:vectorh} become a single \emph{scalar Helmholtz equation}, that is
\begin{equation}\label{eq:helmholtz}
\nabla^2 \varphi (x,y) + \epsilon(x,y) k^2 \varphi(x,y) = 0,
\end{equation}
where $\varphi$ stands for either the $E_z$ or $H_z$ field component.
For the remainder, we shall also suppose that optical geometries have infinite extent in the direction perpendicular to the $(x,y)$ plane and no propagating component in the $z$ direction.
This allows one to ignore the $z$ dependence of the EM field.
We shall also assume that optically thin geometries, such as slab waveguides, can be taken into account via an effective value of the permittivity $\epsilon(x,y)$.
As with \eqref{eq:vectore} and \eqref{eq:vectorh}, the scalar form of the equation means the eigenfunctions of the Helm\-holtz equation are of similar form for both polarizations, with however differing boundary conditions (see for instance section \ref{sec:boundary}).
Let us recall once again that equations \eqref{eq:vectore}, \eqref{eq:vectorh} and \eqref{eq:helmholtz} apply only for constant or piecewise constant refractive index.

\subsubsection{Scattering vs resonances: Quasi-bound states}
\label{sec:qb}
The linear Helmholtz equation \eqref{eq:helmholtz} may be used for two kinds of computations, namely wave scattering and resonances.
As will be discussed in section \ref{sec:scattering}, scattering computations using 2D-GLMT are straightforward and consist in computing the response of a given optical system to an incident excitation. 
The incident and scattered wave are expanded on a suitable basis of the Helmholtz equation (in the case of 2D-GLMT, cylindrical harmonics) and the expansion coefficients of the scattered wave are deduced from the expansion coefficients of the incident excitation.
Thus, wave scattering problems imply solving a inhomogeneous linear system of equations.

On the other hand, resonance computations are more subtle since they are analogous to solving a homogeneous system, in other words finding the non-linear \emph{eigenvalues} of a system of equations.
The associated solutions of the Helmholtz equation are consequently termed \emph{eigenmodes}. 
Computing the eigenmodes usually implies searching for complex $k$ values $(k = k' + ik'')$ satisfying a \emph{resonance condition}. 
Complex $k$ values with $k'' < 0$ result in exponential decay of the total energy of the optical system. According to the harmonic hypothesis $\varphi(\mathbf{r},t) = \varphi(\mathbf{r})e^{-i\omega t}$, the total electromagnetic energy of the system is proportional to
\begin{equation}
|\varphi(\mathbf{r},t)|^2 = |\varphi(\mathbf{r})|^2 e^{- t / \tau},
\end{equation}
where $|\varphi(\mathbf{r})|^2$ is the field profile of the eigenmode and $\tau$ the characteristic decay time (or half-life) of the system energy. It is given by
\begin{equation}
\tau = - \frac{1}{2 c k''}.
\end{equation}
This delay can be readily interpreted as the characteristic residency time of a photon trapped in an optical resonator. 
Since these modes are inherently leaky, they are usually called \emph{meta-stable} or \emph{quasi-bound} (QB) states, by opposition to bound states characteristic of closed quantum systems. 
This analogy stems from the similarity between the Schr\"odinger equation and the Helmholtz equation \cite{Noeckel1997}.

Although useful for modelling active cavities, QB states come with an important difficulty.
Whereas the solutions of \eqref{eq:helmholtz} are assumed to be stationary inside the cavity, resonant wavefunctions of open systems cannot by definition be stationary since their energy must decay in time \cite{Harayama2011}. 
Since $\epsilon(\mathbf{r})$ is real everywhere, the only way to obtain a decaying energy is through the imaginary part of the eigenfrequency $k$.
However, this imaginary part has the net effect of introducing an artificial gain in the whole space domain. \index{Gain}
Consequently, QB states grow exponentially towards infinity, which is not physically realistic \cite{Tuereci2006}.
Strictly speaking, these states are not even regular functions because of this blow-up behaviour \cite{Harayama2011}.
In section \ref{sec:salt}, we introduce a recently formulated theory that overcomes this shortcoming by taking into account the active medium \emph{ab initio}, instead of the \emph{a posteriori} introduction of gain associated with QB states \cite{Tuereci2006}.

\subsection{Optically active media: Steady-state ab initio laser theory (SALT)}
\label{sec:salt}

In the previous section, we have derived the scalar Helmholtz equation \eqref{eq:helmholtz} for 2D geometries. The main hypothesis behind the derivation is the linear response of the optical media. 
In the presence of an active medium, for instance a pumped laser cavity, the derivation is somewhat different.
As discussed in section \ref{sec:qb}, it is still possible however to use the linear Helmholtz equation to model 2D active media using QB states. 
This description comes with two main shortcomings: the exponential growth of QB states outside the resonator as mentioned previously and the conspicuous absence of the gain medium parameters in the linear Helmholtz equation \eqref{eq:helmholtz}.

To circumvent this difficulty, we will make use of a recent formulation called \emph{steady-state ab initio laser theory} (SALT).
The term \emph{ab initio} refers to the fact that SALT only requires the distribution of $\epsilon(\mathbf{r})$ of the passive cavity, or resonator, and a number of parameters describing the gain medium. 
The theory is stationary, meaning it works in the frequency domain, and is intended to bridge the gap between the over-simplified QB states approach described in section \ref{sec:qb} and time-domain simulations using dynamical theories, for instance the Maxwell-Bloch or Schr\"odinger-Bloch theories \cite{Tuereci2006, Harayama2011}.
Partly equivalent formulations devised for the extraction of the lasing thresholds and frequencies of optical cavities are also presented in Refs. \cite{Nojima2005, Smotrova2011, Chang2012}.

SALT was initially developed by H. T\"ureci, A. Stone and B. Collier in 2006, and originally called \emph{ab initio self-consistent laser theory}, or AISC \cite{Tuereci2006}. 
It was this seminal paper that first introduced the constant-flux (CF) states central to the \emph{ab initio} approach. 
The theory was further developed in \cite{Tuereci2007, Tuereci2008, Ge2010th, Ge2010} and renamed SALT in the process.
Further generalization of SALT to complex gain media is to this day an active research topic \cite{Cerjan2015}.
In section \ref{sec:basicsalt}, we highlight the defining features of the theory and present the governing equations for the eigenmodes of a 2D laser cavity near threshold. 
Since the CF states of an array of active cylinders can be readily computed using 2D-GLMT, we discuss them further in section \ref{sec:cf}.

\subsubsection{Basic equations and threshold lasing modes}
\label{sec:basicsalt}

In this section, we explain how to obtain the basic equations of SALT using the electromagnetic theory presented in section \ref{sec:em} and the Maxwell-Bloch equations for a two-level atomic system.
Since the goal is not to present a detailed discussion of SALT but rather to describe the eigenstates that are compatible with 2D-GLMT, we shall only present an abridged derivation of the SALT equations and briefly discuss the physical properties of the eigenstates.
For convenience, only the derivation for a TM polarized wave and a 2D optical geometry is presented.
We refer the interested reader to \cite{Tuereci2006} or \cite{Ge2010th} for a complete theoretical description of SALT.

The first step in the derivation of the SALT equations is to introduce a non-linear polarization term of the following form
\begin{equation}
\mathbf{P}_\mathrm{T} = \chi_e \mathbf{E} + \mathbf{P}_{\mathrm{NL} }, \label{eq:pnlinear}
\end{equation}
in \eqref{eq:nne}, which yields
\begin{equation}\label{eq:nne2}
\nabla \times \nabla \times \mathbf{E} + \frac{\epsilon}{c^2}\frac{\partial^2 \mathbf{E} }{\partial t^2} 
+ \frac{1}{c^2}\frac{\partial^2 \mathbf{P}_{\mathrm{NL}} }{\partial t^2} = 0.
\end{equation}
Under the assumption of a piecewise constant refractive index, and for TM polarization with $E \equiv E_z$, \eqref{eq:nne2} simplifies to
\begin{equation}\label{eq:helmholtzpol}
\nabla^2 E - \frac{\epsilon}{c^2}\frac{\partial^2 E }{\partial t^2} 
- \frac{1}{c^2}\frac{\partial^2 P }{\partial t^2} = 0,
\end{equation}
where $P \equiv ( P_{\mathrm{NL}} )_z$.
Under the rotating wave approximation \cite{Ge2010}, one can expand the electric and polarization fields in a positive frequency part and a negative frequency part, i.e. $E = E^+ + E^-$ and $P = P^+ + P^-$.
This allows to rewrite \eqref{eq:helmholtzpol} only for the positive frequency field component, keeping in mind that the negative frequency component satisfies a similar equation
\begin{equation}\label{eq:helmholtzpol2}
\nabla^2 E^+ - \frac{\epsilon}{c^2}\frac{\partial^2 E^+ }{\partial t^2} 
- \frac{1}{c^2}\frac{\partial^2 P^+ }{\partial t^2} = 0.
\end{equation}
The SALT theory is based on the Maxwell-Bloch equations for a two-level atomic system \cite{Tuereci2006}:
\begin{gather}
\frac{\partial P^+}{\partial t} = - (i \omega_a + \gamma_\perp) P^+ + \frac{g^2}{i\hbar} E^+ \mathcal{D}, \label{eq:dpol} \\
\frac{\partial \mathcal{D}}{\partial t} = 
\gamma_\parallel ( \mathcal{D}_0( \mathbf{r}) - \mathcal{D} ) - \frac{2}{i\hbar} [ E^+ (P^+)^* +  P^+ (E^+)^* ]. \label{eq:dinv}
\end{gather}
These equations relate the electric and polarization fields with the spatially varying population inversion $\mathcal{D}(\mathbf{r})$ and the pump profile $\mathcal{D}_0( \mathbf{r})$. The gain central frequency $\omega_a$ and linewidth (or polarization relaxation rate) $\gamma_\perp$ appear in \eqref{eq:dpol}. \index{Gain!center frequency} \index{Gain!width}
Other constants are the population relaxation rate $\gamma_\parallel$ and the dipole matrix element $g$. For convenience, we renormalize the population inversion
\begin{equation}
D \equiv \frac{g^2 \mathcal{D}}{ \gamma_\perp \hbar},
\end{equation}
so that $D$ is dimensionless. This results in the following Maxwell-Bloch equations
\begin{gather}
\frac{\partial P^+}{\partial t} = - (i \omega_a + \gamma_\perp) P^+ -i \gamma_\perp E^+ D, \label{eq:dpol2} \\
\frac{1}{\gamma_\parallel}\frac{\partial D}{\partial t} = 
(D_0( \mathbf{r}) - D ) +i \kappa [ E^+ (P^+)^* +  P^+ (E^+)^* ], \label{eq:dinv2}
\end{gather}
where the coupling constant $\kappa$ is defined as
\begin{equation}
\kappa \equiv \frac{2 g^2}{\gamma_\parallel \gamma_\perp \hbar^2}.
\end{equation}
This coupling coefficient governs the strength of the spatial-hole burning effect, that is interaction between lasing modes above threshold.

The main hypothesis of SALT resides in the so-called \emph{stationary inversion approximation}, which allows to recast the Maxwell-Bloch equations in a numerically tractable form.
In short, this approximation implies that the population inversion of the two-level system remains constant in time ($\frac{\partial D}{\partial t} = 0$). 
This approximation is valid in the single-mode regime, as well as in the multi-mode regime under the condition $\gamma_\perp \gg \gamma_\parallel$, if the characteristic time scale of the inversion dynamics is larger than that of the polarization dynamics. 
Furthermore, in the single-mode regime or when the electromagnetic fields are small with respect to the coupling coefficient $\kappa$, the hole-burning effects can be neglected.
In this case, a multiperiodic Ansatz yields a set of \emph{threshold lasing modes}, or TLMs:
\begin{subequations}
	\begin{gather}
	E^+ (\mathbf{r},t) = \sum_\mu \varphi_\mu(\mathbf{r}) e^{-ick_\mu t}, \label{eq:tlm1} \\
	P^+ (\mathbf{r},t) = \sum_\mu p_\mu(\mathbf{r}) e^{-ick_\mu t}. \label{eq:tlm2}
	\end{gather}
\end{subequations}
This set of modes is governed by the following equation
\begin{equation}\label{eq:tlm}
\bigg\lbrace \nabla^2 + \left[ \epsilon(\mathbf{r}) + \dfrac{\gamma_a D_0^\mu F(\mathbf{r})}{k_\mu - k_a + i \gamma_a} \right] k_\mu^2 \bigg\rbrace \varphi_\mu (\mathbf{r}) = 0,
\end{equation}
where $k_a = \omega_a/c$ is the gain centre frequency (we use frequency and wavenumber interchangeably), $\gamma_a = \gamma_\perp /c$ is the gain width, $D_0^\mu$ is associated to the ``pump strength'' and $F(\mathbf{r})$ is the spatial pump profile. 
In the absence of pumping ($F(\mathbf{r}) = 0$), for instance outside the laser resonator, the modes satisfy a linear Helmholtz equation \eqref{eq:helmholtz} with real frequency $k$. 
This is a more realistic description of the lasing modes than the QB states approach since the gain medium is introduced as an effective complex-valued permittivity inside the resonator instead of an artificial imaginary part of the wavenumber \cite{Tuereci2006}.

The determination of TLMs can be achieved using various numerical schemes including the Lorenz-Mie approach described in section \ref{sec:glmt} or the finite element method described in \cite{Liertzer2012}.
For a given combination of $\epsilon(\mathbf{r})$, $F(\mathbf{r})$, $k_a$ and $\gamma_a$, the problem consists in finding the value of the lasing frequencies $k_\mu$, thresholds $D_0^\mu$ and the field distributions $\varphi_\mu$. 
However, for a given choice of exterior real frequency $k$, the value of $D_0^\mu$ is in general complex.
Consequently, the TLM must satisfy an additional reality condition on the threshold value, since a complex-valued $D_0^\mu$ does not correspond to a physically realistic lasing mode.
This means that one must map the evolution of $D_0^\mu$ as a function of $k$, and when the former crosses the real axis at $k=k_\mu$, obtain a pair of \emph{real} numbers $(k_\mu, D_0^\mu)$ defining the TLM lasing frequency and lasing threshold, respectively.
The first lasing mode is therefore the TLM with the smallest threshold $D_0^\mu$ \cite{Ge2010}.

\subsubsection{Constant-flux states}
\label{sec:cf}

One of the defining features of SALT is the introduction of new eigenstates, such as the TLMs described by \eqref{eq:tlm}. The theory also introduces another kind of eigenstate called a constant-flux (CF) state \cite{Tuereci2006}.
CF states are useful because they are straightforward to compute and can be used as a basis to expand TLMs \cite{Ge2010}.
They are parametrized by real wavenumbers outside the laser resonator, and are thus physically meaningful when compared to QB states. 
The basis of CF states satisfy the following modified Helmholtz equation
\begin{subequations}
\begin{align}
[\nabla^2 + \epsilon(\mathbf{r}) K^2_\mu(k)]\varphi_\mu &= 0, \qquad \mathbf{r} \in C, \label{eq:cfa} \\
[\nabla^2 + \epsilon(\mathbf{r}) k^2]\varphi_\mu &= 0, \qquad \mathbf{r} \notin C,
\end{align}\label{eq:helmholtzcf}
\end{subequations}
where $C$ is the \emph{cavity region}, defined as the union of all optically active regions. The eigenvalues $K_\mu$ are complex and depend on the \emph{exterior frequency} $k$, which is always real. This formulation ensures that the total electromagnetic flux outside the cavity is conserved \cite{Tuereci2006}.

In the special case of a uniform pumping inside the cavity region, i.e. $F(\mathbf{r}) = 1$ for $\mathbf{r} \in C$ and $F(\mathbf{r})=0$ for $\mathbf{r} \notin C$, and a uniform refractive index distribution $\epsilon(\mathbf{r}) = \epsilon_c $ for $\mathbf{r} \in C$, the TLMs can be trivially expanded in the basis of CF states. \index{Pumping!uniform}
By comparing \eqref{eq:tlm} and \eqref{eq:cfa}, one obtains the relation
\begin{equation}\label{eq:relation}
\dfrac{\gamma_a D_0^\mu}{k - k_a + i \gamma_a} = \epsilon_c \left( \frac{K_\mu^2}{k^2} - 1 \right).
\end{equation}
In other terms, if the active medium is uniformly pumped, one can simply compute the CF states of the geometry, given the value of $\epsilon_c$, and extract the complex values $D_0^\mu$ of the associated TLMs using \eqref{eq:relation}.
Moreover, in the simplest possible approximation, CF states may be regarded as the \emph{actual} lasing modes, for instance in the case of a gain transition frequency close to the CF state eigenfrequency \cite{Tuereci2006}.

\subsection{Circular cavity: QB vs CF states}

This section completes the description of QB and CF states by giving their explicit expressions for the simplest 2D geometry for which the solution is analytic: a single homogeneous dielectric cylinder, or circular cavity.
Since this tutorial is mostly concerned with coupled cylinder geometries, it is also useful to present the closed-form expressions for the eigenmodes of a single, uncoupled cylinder, which is a limiting case of the Lorenz-Mie theory presented in section \ref{sec:glmt}.

Consider a dielectric circle of refractive index $n_c=\sqrt{\epsilon_c}$ and radius $r$ placed in a medium of refractive index $n_0$. The cavity is centred at the origin of a ``global'' cylindrical coordinate system. Under these conditions, the QB states of the circular cavity, solutions of Eq. \eqref{eq:helmholtz}, are \cite{Noeckel1997}
\begin{equation}
\varphi_l(\rho,\theta) = 
\begin{cases}
A_l J_l (n_c k \rho), \qquad &\rho < r, \\
B_l \h_l (n_0 k \rho), \qquad &\rho > r,
\end{cases}\label{eq:circleqb}
\end{equation}
where $l$ is the angular quantum number of the solution, $J_l$ is a Bessel function of the first kind and $\h_l = J_l + i Y_l$ is a Hankel function of the first kind.  Analogously, the CF states of the circular cavity, solutions of Eq. \eqref{eq:helmholtzcf}, are \cite{Tuereci2006}
\begin{equation}
\varphi_l(\rho,\theta) = 
\begin{cases}
A_l J_l (n_c K(k) \rho), \qquad &\rho < r, \\
B_l \h_l (n_0 k \rho), \qquad &\rho > r.
\end{cases}\label{eq:circlecf}
\end{equation}
One notes that the CF eigenvalue $K(k)$ explicitly appears in the solution for $\rho < r$.

There exists an infinite set of QB and CF states for any cavity. In the case of the circular cavity, both types of states can be characterized by a pair of quantum numbers $(l,j)$, corresponding to the number of angular and radial lobes of the eigenmode, respectively. Moreover, the expressions given by Eqs. \eqref{eq:circleqb} and \eqref{eq:circlecf} satisfy an outgoing boundary condition outside the resonator, sometimes called \emph{Sommerfeld radiation condition}.
This condition ensures a net flow of the electromagnetic energy of the eigenstate towards infinity, and is mathematically given by
\begin{equation}\label{eq:sommerfeld}
\lim_{\rho \rightarrow \infty} \varphi(\rho,\theta) = \varphi(\theta) \frac{e^{ik \rho} }{\sqrt{k \rho}}.
\end{equation}
In order to determine the eigenvalues of QB states, one must find non-trivial solutions to a homogeneous linear system relating the unknown coefficients $A_l$ and $B_l$. This linear system, obtained via the application of electromagnetic boundary conditions (see section \ref{sec:boundary})
at $\rho=r$ is given by
\begin{equation}\label{eq:charqb}
\begin{pmatrix}
J_l (n_c k r) & - \h_l (n_0 k r) \\
\varsigma_n n_c J_l'(n_c kr) & - \varsigma_0 n_0 \hp_l (n_0 k r)
\end{pmatrix} 
\begin{pmatrix}
A_l \\ B_l
\end{pmatrix} = 0,
\end{equation}
where $\varsigma_i= 1$ $(1/n_i^2)$ for a TM (TE) polarized wave. These factors account for the polarization of the mode.
In essence, one must find the complex eigenvalues $k=k_{\mathrm{QB}}$ such that the determinant of the coefficient matrix in Eq. \eqref{eq:charqb} is zero. Once this is done, the $A_l$ and $B_l$ coefficients are readily obtained from the linear system. A similar system of equations can be obtained for the CF states of the circular cavity
\begin{equation}\label{eq:charcf}
\begin{pmatrix}
J_l (n_c K r) & - \h_l (n_0 k r) \\
\varsigma_n n_c K J_l'(n_c Kr) & - \varsigma_0 n_0 k \hp_l (n_0 k r)
\end{pmatrix} 
\begin{pmatrix}
A_l \\ B_l
\end{pmatrix} = 0.
\end{equation}
In this case, one must find eigenvalues $K_\mu(k)$ for a fixed exterior frequency $k$.

\begin{figure*}
\centering
\includegraphics[]{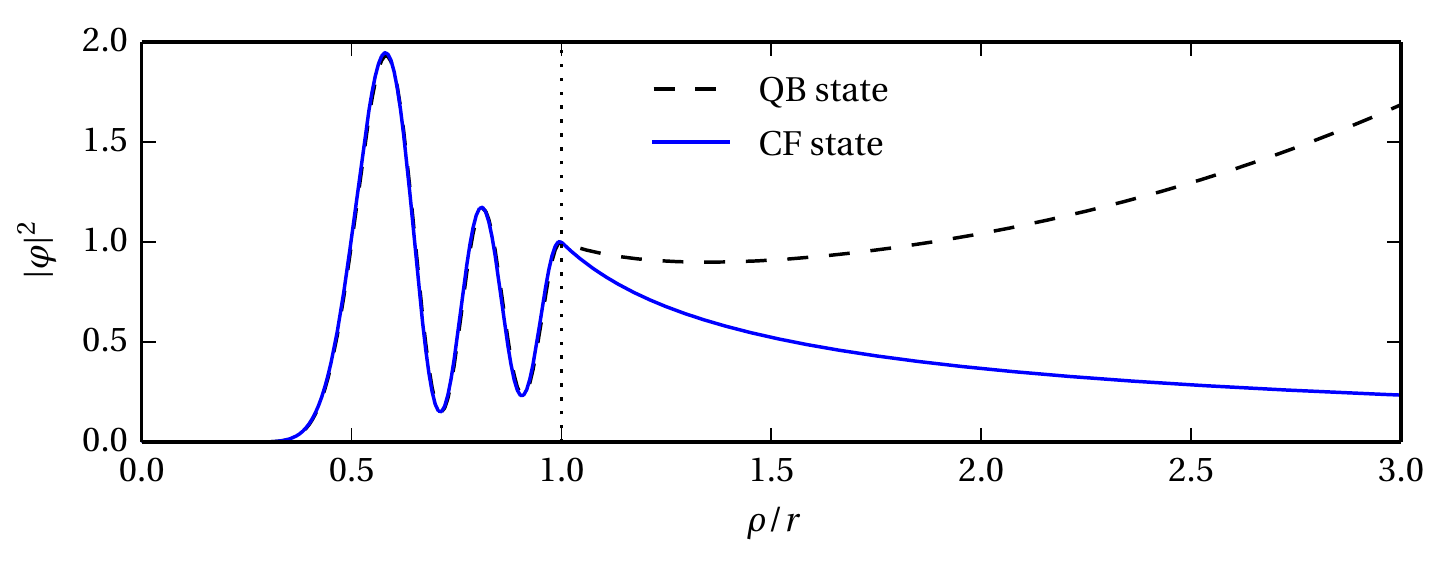}
\caption[Comparison of the field profile of a QB and a CF state of a circular cavity]{Comparison of the field profile of a QB and a CF state of a circular cavity of refractive index $n_c=1.5$ embedded in a medium of refractive index $n_0 = 1$. (TM polarization, arbitrary normalization). The eigenmodes are characterized by the quantum numbers $(l,j) = (10,3)$. The QB state eigenvalue is $k_{\mathrm{QB}} = 13.521 - 0.442i$. The CF state eigenvalue is $K_\mu = 13.558 - 0.440i$ (exterior frequency $k=13.52$). Reproduced from \cite{Gagnon2014th}.}\label{fig:qbcf}
\end{figure*}

For comparison, a QB state of the circular cavity and the associated CF state are plotted in Fig. \ref{fig:qbcf}. Both modes are characterized by the pair of quantum numbers $(l,j) = (10,3)$. 
Since we chose the exterior frequency of the CF state close to the real part $k'_{\mathrm{QB}}$ of the QB eigenvalue, the field profile inside the cavity $(\rho < r)$ is similar for both eigenstates. 
However, one clearly sees that the CF state remains bounded outside the cavity, whereas the QB state begins to grow exponentially roughly beyond $\rho = 1.5 r$.
This behaviour is consistent with the Sommerfeld radiation condition.
If one injects a complex $k$ value in Eq. \eqref{eq:sommerfeld} with $k'' < 0$, the solution clearly blows up at infinity.
In contrast, CF states are always characterized by real wavenumbers outside the cavity region, which ensures that the solution remains bounded.
This example of the circular cavity serves two purposes. Firstly, it shows that QB and CF states can be uniquely mapped onto each other for the corresponding quantum numbers.
Actually, this mapping is also possible in the case of more complex cavity geometries \cite{Tuereci2006}, a prime example of which is random lasers, discussed for instance in \cite{Tuereci2008, Tuereci2009, Andreasen2011}.
Secondly, it highlights the unphysical behaviour of the usual QB states at infinity, a behaviour which can be corrected using the formulation of CF states.

To conclude this section on SALT, the main features of the three kinds of eigenstates (QB states, CF states and TLMs) are summarised in Table \ref{tab:eigenstates}.

% % % A FIGURE AND A TABLE TO COMPARE QB and CF STATES % % % %
%\begin{landscape}

\begin{table*}
\caption[Comparison of the three kinds of eigenstates used in this thesis]{Comparison of the three kinds of eigenstates used in this tutorial. Outside the cavity region, all eigenstates are governed by the linear Helmholtz equation \eqref{eq:helmholtz}. The defining features listed in this table assume that the refractive index distribution $\epsilon(\mathbf{r})$ is a real and piecewise constant function.}\label{tab:eigenstates}
\begin{center}
\bgroup
\def\arraystretch{1.5}
\begin{tabular}{ccp{6.0cm}p{5cm}}
Name	& Acronym	& Governing equation \emph{inside} cavity region & Defining features \\ \hline
Quasi-bound state	& QB	& $[\nabla^2 + \epsilon(\mathbf{r}) k^2]\varphi_\mu(\mathbf{r}) = 0$ & Grows exponentially at infinity; $k$ complex \\
Constant-flux state & CF	& $[\nabla^2 + \epsilon(\mathbf{r}) K_\mu^2(k)]\varphi_\mu(\mathbf{r}) = 0$ & Remains bounded at infinity; $K_\mu$ complex, $k$ real \\
Threshold lasing mode & TLM	& $\bigg\lbrace \nabla^2 + \left[ \epsilon(\mathbf{r}) + \dfrac{\gamma_a D_0^\mu F(\mathbf{r})}{k_\mu - k_a + i \gamma_a} \right] k_\mu^2 \bigg\rbrace \varphi_\mu = 0$ & Includes a description of the pump and gain medium; $(k_\mu, D_0^\mu)$ real \\ \hline
\end{tabular}
\egroup
\end{center}
\end{table*}
%\end{landscape}

\section{2D Generalized Lorenz-Mie Theory}
\label{sec:glmt}

We now describe in some details the theoretical framework which we refer to as 2D-GLMT.
According to Gouesbet and Lock \cite{Gouesbet2013}, the expression ``generalized Lorenz-Mie theories''
(GLMTs) is used to generically denote a set of light-scattering theories describing the interaction between an incident electromagnetic beam and a discrete scattering particle (or array of particles) provided that the scatterers possess enough symmetry, allowing to solve the problem using the method of separation of variables. 
The approach has been referred to by a number of names in the past, including ``multipole method'' or ``fast multipole method'' in Refs. \cite{Andreasen2011, Kuhlmey2002, Vukovic2010, White2002}. 
Historically, one should mention that neither Ludvig Lorenz nor Gustav Mie were involved in the solution of Maxwell's equations of coupled cylinders. The solution of Maxwell's equations in cylindrical geometries is best associated to Lord Rayleigh \cite{LordRayleigh1918}, and some authors prefer the term ``Rayleigh scattering'' for cylinders instead of ``Lorenz-Mie scattering'' or ``Mie scattering'' which usually deals with spheres \cite{Gouesbet2013}. 
We will avoid the term ``Rayleigh scattering'' since it most often refers to scattering by particles much smaller than the wavelength of the incident light, an approximation not used in the present treatment.

This section presents a ``reference implementation'' of the 2D-GLMT and its application to electromagnetic modelling of two-dimensional arrays of cylinders. 
The derivations and notation used follow references \cite{Elsherbeni1992, Felbacq1994, Nojima2005, Oguzer1995}. 
Similar derivations are also found in a variety of other published works including \cite{Andreasen2011, Natarov2014}.

This section is organized as follows. The first sub-sections (sections \ref{sec:graf} to \ref{sec:boundary})  are concerned with the basic theoretical treatment of 2D-GLMT. In section \ref{sec:graf}, we state Graf's addition theorem for cylindrical functions since it is at the heart of 2D-GLMT.  Section \ref{sec:basic} presents the basic equations and the main hypothesis of 2D-GLMT, the expansion of the electromagnetic fields in a basis of cylindrical waves. To isolate the unknown expansion coefficients, electromagnetic boundary conditions must then be enforced as described in section \ref{sec:boundary}.

The next two sub-sections deal with two classes of computations enabled by the use of 2D-GLMT. 
In section \ref{sec:scattering}, we describe \emph{scattering computations}, for which
2D-GLMT is an ideal tool to obtain the field produced by a finite array of cylindrical scatterers. We show how this formalism is compatible not only with the scattering of plane waves by cylinders, but also of non-paraxial Gaussian beams via the complex-source beam approach.
\emph{Eigenmode calculations} follow in section \ref{sec:eigenmodes}, where lasing modes of an array of cylinders are presented.
More specifically, we will be interested not only in the computation of the usual quasi-bound (QB) states,
but also of the constant-flux (CF) states, central to the SALT theory.
An equivalent method for computing modes of random lasers has been summarily described in \cite{Andreasen2011} under the name ``multipole method''.

\subsection{Graf's addition theorem}
\label{sec:graf} 
\begin{figure}
 \centering
 \includegraphics[width=\columnwidth]{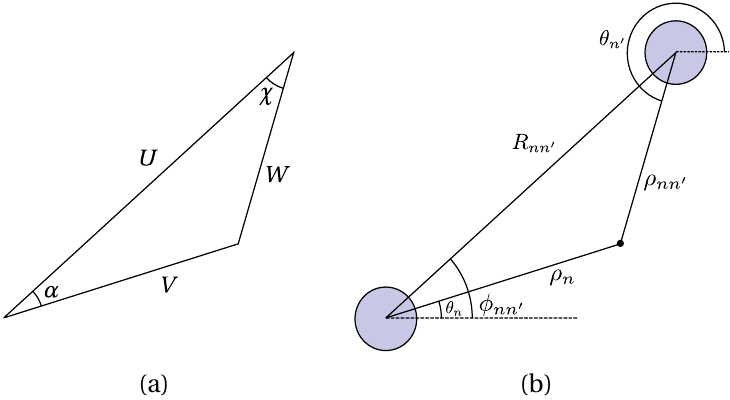}
\caption[Graf's addition theorem]{Graf's addition theorem. (a) Triangle interpretation of the theorem. (b) Application of the theorem to Eq. \eqref{eq:graf}.}\label{fig:graf_theorem}
\end{figure}

\index{Graf's addition theorem}
We state Graf's addition theorem  \cite[Eq.~9.1.79]{Abramowitz1970} since it is central to the derivation of the main equations of 2D-GLMT. This addition formula allows one to displace Bessel functions from one cylindrical system of coordinates into another.
Let $\mathcal{F}$ denote $J, Y, \h, \hh$ or any linear combination of these functions. The following identity holds
\begin{equation}
\mathcal{F}_\nu (W) e^{i\nu \chi} = \sum_{m=-\infty}^{\infty} \mathcal{F}_{\nu + m} (U) J_m(V) e^{im\alpha},
\end{equation}
where $|V e^{\pm i \alpha}| < |U|$ and
\begin{equation}
\begin{aligned}
&W^2 = U^2 + V^2 - 2UV \cos \alpha, \\
&U - V \cos \alpha = W \cos \chi, \\ 
&V\sin \alpha = W \sin \chi.
\end{aligned}\label{eq:grafcv}
\end{equation}
The branches must be chosen such that $W \rightarrow U$ and $\chi \rightarrow 0$ as $V \rightarrow 0$. The restriction $|V e^{\pm i \alpha}| < |U|$ is unnecessary if $\mathcal{F} = J$ because the Bessel function of the first kind and integer order is a bounded entire function. If $U,V,W$ are real numbers, they can be interpreted as edges of a triangle as shown in Fig. \ref{fig:graf_theorem}a. The theorem holds for complex-valued arguments also.
%\begin{figure*}
%\end{figure*}

\subsection{Basic equations}
\label{sec:basic}
Consider an array of $N$ cylindrical scatterers of radii $r_n$ and relative permittivity $\epsilon_n$. Let also $\mathbf{r}_n = (\rho_n,\theta_n)$ be the cylindrical coordinate system local to the $n^\mathrm{th}$ scatterer, whose center is located at $\mathbf{R}_n = (X_n, Y_n)$. 
We suppose that every cylinder (or hole) is infinite along the axial $z$ direction. The field outside the cylinders satisfies the following Helmholtz equation
\begin{subequations}
\begin{equation}
[\nabla^2 + k_0^2]\varphi(x,y) = 0,
\end{equation}
with $k_0 = \sqrt{\epsilon_0} k$, whereas the field inside the $n^\mathrm{th}$ scatterer satisfies the following Helmholtz equation 
\begin{equation}
[\nabla^2 + k_n^2]\varphi(\rho_n,\theta_n) = 0, \qquad \rho_n < r_n.
\end{equation}
\end{subequations}
For generality, we do not suppose any relation between $k_n$ and $\epsilon_n$ at this point. This will allow the derivations to be compatible both with passive and active media, as described in section \ref{sec:salt}.

The central hypothesis of 2D-GLMT is that the field exterior to the scatterers $\varphi^E$ can be written as the superposition of an arbitrary incident beam and the sum of the field scattered by each individual scatterer. The fields can be expanded in a basis of cylindrical functions, that is
\begin{equation}
\varphi^E (\mathbf{r}) = \varphi_i(\mathbf{r}) + \varphi_s(\mathbf{r}),
\end{equation}
with 
\begin{subequations}\label{eq:field}
\begin{gather}
\varphi_i(\mathbf{r}) = \sum_{l=-\infty}^{\infty} a^0_{nl} J_l(k_0 \rho_n)e^{il\theta_n}, \\ 
\varphi_s(\mathbf{r}) = \sum_{n=1}^N \sum_{l'=-\infty}^{\infty} b_{nl'} \h_{l'} (k_0 \rho_{n}) e^{il'\theta_n}.\label{eq:scat}
\end{gather}
\end{subequations}
The coefficients $ \lbrace{ a^0_{nl} \rbrace}$ are the \emph{beam-shape coefficients}, used to parametrize the incident field in the frame of reference of the $n^\mathrm{th}$ scatterer.

Inside the $n^\mathrm{th}$ scatterer, the field is similarly represented as 
\begin{equation}\label{eq:infield}
\varphi^{I}_n(\mathbf{r}) = \sum_{l=-\infty}^{\infty} c_{nl} J_l(k_n \rho_n)e^{il\theta_n}.
\end{equation}
In order to apply electromagnetic boundary conditions at the interface of the $n^\mathrm{th}$ scatterer, one must find an expression for $\varphi^E(\mathbf{r})$ containing only cylindrical harmonics centred on the $n^\mathrm{th}$ scatterer, that is
\begin{equation}\label{eq:ab}
\varphi^E_n(\mathbf{r}) = \sum_{l=-\infty}^{\infty} \left[ a_{nl} J_{l}(k_0 \rho_n) + b_{nl} \h_{l}(k_0 \rho_n) \right] e^{il\theta_n}.
\end{equation}
This can be achieved via the application of Graf's addition theorem for cylindrical functions, performing a translation from the frame of reference of scatterer $n'$ to the frame of reference of scatterer $n$. We apply Graf's addition theorem with an appropriate change of variables (see Fig. \ref{fig:graf_theorem}b for the definition of angles and edges)
\begin{equation}
U = k_0 R_{nn'}, V = k_0 \rho_n, W = k_0 \rho_{n'}.
\end{equation}
The theorem states that
\begin{widetext}
\begin{equation}\label{eq:graf}
 \h_{l'} (k_0 \rho_{n'}) e^{il' \theta_{n'}} = \sum_{l=-\infty}^{\infty}e^{i(l'-l) \phi_{nn'}} \h_{l-l'}(k_0R_{nn'}) J_l (k_0 \rho_n) e^{il\theta_n},
\end{equation}
where $R_{nn'}$ is the center-to-center distance between scatterers $n$ and $n'$ and $\phi_{nn'}$ is the angular position of scatterer $n'$ in the frame of reference of scatterer $n$. Substituting \eqref{eq:graf} in \eqref{eq:field} yields
\begin{equation}\label{eq:graf2}
\begin{aligned}
\varphi^E_n(\mathbf{r}) & = \sum_{l=-\infty}^{\infty} a^0_{nl} J_l(k_0 \rho_n)e^{il\theta_n} + \sum_{l=-\infty}^{\infty} b_{nl} \h_{l} (k_0 \rho_{n}) e^{il\theta_n} \\
& + \sum_{l=-\infty}^{\infty} \sum_{n' \neq n} \sum_{l'=-\infty}^{\infty} b_{n'l'} e^{i(l'-l) \phi_{nn'}} \h_{l-l'}(k_0 R_{nn'}) J_l(k_0 \rho_n)e^{il\theta_n}.
\end{aligned}
\end{equation}
The comparison of \eqref{eq:ab} with \eqref{eq:graf2} results in the following relation between the $\lbrace a_{nl} \rbrace$ and $\lbrace b_{nl} \rbrace$ coefficients
\begin{equation}\label{eq:anl}
a_{nl} = a_{nl}^0 + \sum_{n' \neq n} \sum_{l'=-\infty}^{\infty}e^{i(l'-l) \phi_{nn'}} \h_{l-l'}(k_0R_{nn'}) b_{n'l'}.
\end{equation}
\end{widetext}

At this point, we should stress that 2D-GLMT is not restricted to disconnected cylinders embedded in an infinite medium.
Cylinders embedded in other cylinders are also within the reach of the method, provided appropriate expressions of the field are used.
The surrounding cylinder may represent the jacket of an optical fibre, for instance, and the leaky modes of this geometry are readily computed using a slightly modified version of 2D-GLMT, as detailed in Refs. \cite{White2002, Kuhlmey2002}.
For the remainder, however, we shall restrict the discussion to disconnected cylinders with the field in the surrounding medium given by \eqref{eq:graf2}.

\subsection{Electromagnetic boundary conditions and characteristic equation}
\label{sec:boundary}

Relation \eqref{eq:anl} is one of the central results of 2D-GLMT.
It accounts for the mutual influence of all individual scatterers.
In order to obtain the $\lbrace a_{nl} \rbrace$,  $\lbrace b_{nl} \rbrace$ and  $\lbrace c_{nl} \rbrace$ for a given set of beam-shape coefficients $\lbrace a_{nl}^0 \rbrace$, one must apply electromagnetic boundary conditions to \eqref{eq:infield} and \eqref{eq:ab} at $\rho_n = r_n$.
For a TM polarized wave, the first condition is the continuity of $E_z$ across the cylinder interface at $\rho_n = r_n$.
The second condition is the continuity of the component of $\mathbf{H}$ parallel to the interface ($H_{\theta_n}$ component).
From \eqref{eq:ne}, one obtains
\begin{equation}\label{eq:hcont}
H_{\theta_n} = -\frac{i}{k} [\nabla \times \mathbf{E}]_{\theta_n} 
= -\frac{i}{k} \frac{\partial E_z}{\partial \rho_n}.
\end{equation}
For a TE polarized wave, the first condition is rather the continuity of $H_z$ across the cylinder interface at $\rho_n = r_n$, and the second condition is the continuity of the component of $\mathbf{E}$ parallel to the interface ($E_{\theta_n}$ component).
From \eqref{eq:nh}, one obtains
\begin{equation}\label{eq:econt}
E_{\theta_n} = \frac{i}{\epsilon k} [\nabla \times \mathbf{H}]_{\theta_n} 
= \frac{i}{\epsilon k} \frac{\partial H_z}{\partial \rho_n}.
\end{equation}
The four conditions of continuity (two for each polarization) can be rewritten as
\begin{subequations}\label{eq:bcond}
\begin{align}
\varphi^I_n (r_n) &= \varphi^E_n (r_n) \label{eq:psicont}, \\
\varsigma_{n} \dfrac{\partial \varphi^I_n }{\partial \rho_n} \bigg|_{r_n} &= \varsigma_0 \dfrac{\partial \varphi^E_n }{\partial \rho_n} \bigg|_{r_n}.\label{eq:dpsicont}
\end{align}
\end{subequations}
Again, the $\varsigma_i$ factors account for polarization. Specifically, we have $\varsigma_i= 1$ $(1/\epsilon_i)$ for a TM (TE) polarized wave. Thus, applying this boundary condition to \eqref{eq:infield} and \eqref{eq:ab}, one obtains
\begin{subequations}
\begin{gather}
c_{nl} J_l(k_n r_n) = a_{nl} J_l(k_0 r_n) + b_{nl} \h_l (k_0 r_n), \label{eq:cb} \\ 
c_{nl} \varsigma_n k_n J_l'(k_n r_n) = a_{nl} \varsigma_0 k_0 J_l'(k_0 r_n) + b_{nl} \varsigma_0 k_0 \hp_l (k_0 r_n).
\end{gather}
\end{subequations}
Eliminating $c_{nl}$, we obtain the relation $b_{nl}= a_{nl} s_{nl}$, with
\begin{equation}\label{eq:snl}
s_{nl}(k_0, k_n) = -\dfrac{J_l'(k_0 r_n) - \Gamma_{nl} J_l (k_0 r_n)}{\hp_l(k_0 r_n) - \Gamma_{nl} \h_l (k_0 r_n)},
\end{equation}
where
\begin{equation}
\Gamma_{nl} = \xi_{n0} \dfrac{k_n J_l'(k_n r_n)}{k_0 J_l(k_n r_n)},
\end{equation}
and
\begin{equation}
\xi_{ij} = 1 \qquad \left( \frac{\epsilon_j}{\epsilon_i} \right),
\end{equation}
for a TM (TE) polarized wave.

Substituting $b_{nl}= a_{nl} s_{nl}$ in (\ref{eq:anl}) yields
\begin{equation}\label{eq:bnl}
b_{nl} - {s_{nl}}\sum_{n' \neq n} \sum_{l'=-\infty}^{\infty}e^{i(l'-l) \phi_{nn'}} \h_{l-l'}(k_0R_{nn'})b_{n'l'} = {s_{nl}} a_{nl}^0.
\end{equation}
This relation between the $\lbrace a_{nl}^0 \rbrace$ and $\lbrace b_{nl} \rbrace$ coefficients can be rewritten in matrix form as
\begin{equation}\label{eq:system}
\mathbf{T} \mathbf{b} = \mathbf{a}_0,
\end{equation}
with
\begin{equation}\label{eq:tmatrix}
\mathbf{T}_{nn'}^{ll'} = \delta_{nn'}\delta_{ll'} - (1 - \delta_{nn'}) e^{i(l'-l) \phi_{nn'}} \h_{l-l'}(k_0R_{nn'})s_{nl},
\end{equation}
and
\begin{equation}
\mathbf{a}_0 = \bigg\lbrace s_{nl} a^0_{nl} \bigg\rbrace \qquad, \qquad 
\mathbf{b} = \bigg\lbrace b_{nl} \bigg\rbrace.
\end{equation}
The \emph{transfer matrix} $\mathbf{T}$ is typically constructed by truncating the series expansions to order $l_{\mathrm{max}}$. It is composed of $N \times N$ blocks of dimension $2l_{\mathrm{max}} + 1$, where $l_{\mathrm{max}}$ is the truncation order. 
To ensure an adequate representation of the field, the truncation order is often chosen as a linear multiple of the ratio between the characteristic dimension of the problem, for instance the radius of the larger cylinder comprised in the array, and the wavelength.
Irrespective of the value of $l_{max}$, the size of the transfer matrix scales like $N^2$, where $N$ is the number of cylinders considered.
This suggests that 2D-GLMT is best suited to cylinder dimensions of the order of the operating wavelength, in which case it provides fast and accurate results without the need for a spatial discretization of the problem.
Otherwise, in the case of cylinders much larger than the operating wavelength, and/or a very large number of scatters, the memory and processor requirements of the method may become impractical.
In that case, alternative methods may be more appropriate than 2D-GLMT, for instance the finite element method \cite{Fujii2012, Liertzer2012}. 
\begin{figure}
	\centering
	\includegraphics[width=0.25\textwidth]{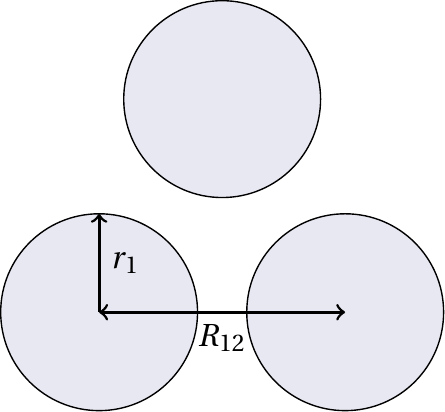}
	\caption{Example scatterer array consisting of three identical coupled cylinders with permittivity $\epsilon_c = 4$, arranged on the vertices of an equilateral triangle ($R_{12} = 2.5 r_1$). The transfer matrix of this arrangement is shown in Fig. \ref{fig:tmatrix}. Reproduced from \cite{Gagnon2014th}.}\label{fig:array}
\end{figure}

\begin{figure*}
\centering
	\includegraphics[width=0.9\textwidth]{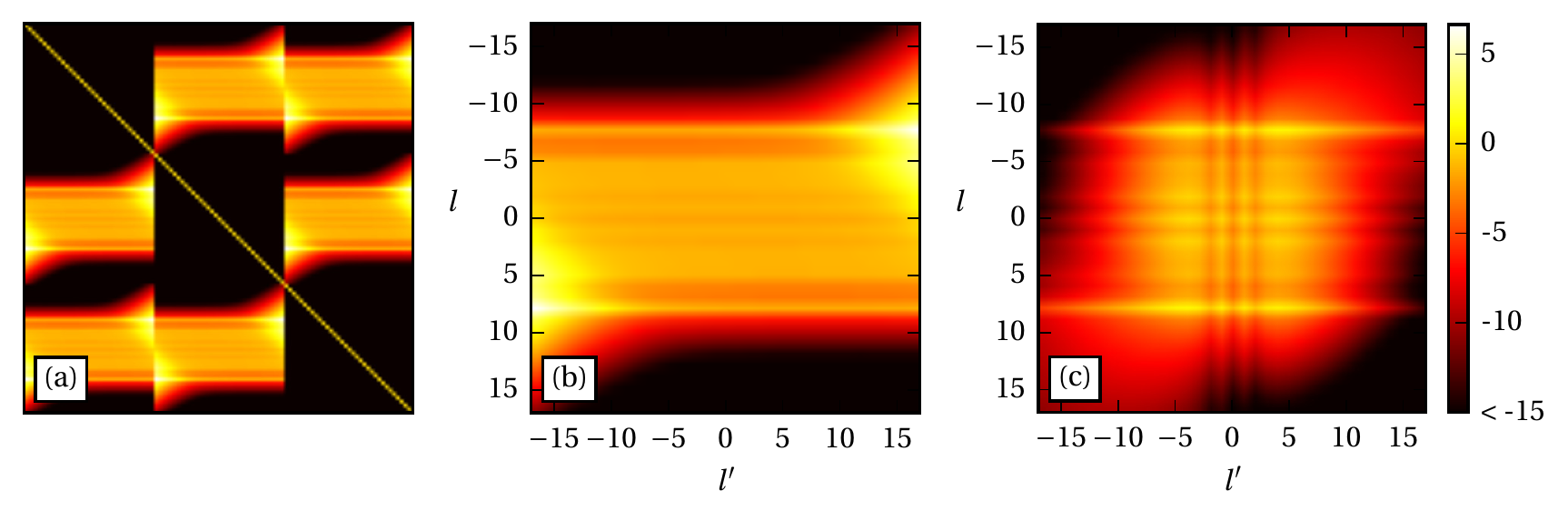}
	\caption{Transfer matrix for the triangle geometry shown in Fig. \ref{fig:array} for $kr_1 = 5.3779$. The truncation order of the matrix is chosen as $l_{\mathrm{max}} = \mathrm{int} [ 3 k r_{1} ] + 1 = 17$. (a) Magnitude of $\log|\mathbf{T}|$ (b) Sub-matrix view of one off-diagonal block of $\log|\mathbf{T}|$. (c) Sub-matrix view of one off-diagonal block recast to a Fredholm form of the second-kind.}\label{fig:tmatrix}
\end{figure*}

As a representative example, the transfer matrix for an array of three identical ($\epsilon_c = 4$) coupled dielectric cylinders arranged on the vertices of an equilateral triangle (Fig. \ref{fig:array}) is shown in Fig. \ref{fig:tmatrix}. This matrix is clearly composed of $3 \times 3$ blocks of dimension $2l_{\mathrm{max}} + 1$, where the prescription $l_{\mathrm{max}} = 3 k r_{1}$ sets the truncation order \cite{Elsherbeni1992, Felbacq1994}.

\subsection{The characteristic equation as a Fredholm form of the second-kind}
\label{sec:fredholm}

Although 2D-GLMT in the form presented in the previous section has been widely used for two decades, it comes with an important shortcoming.
Specifically, the underlying matrix equation \eqref{eq:system} cannot be truncated at an arbitrary order to provide convergence.
In this section, we isolate the source of this difficulty and show how it can be completely alleviated using a straightforward procedure.
This has been acknowledged in recent publications including \cite{Smotrova2006}, and generally solved in \cite{Natarov2014}.

As can be demonstrated by using the asymptotic form of cylindrical functions for large positive order \cite{Abramowitz1970}, for large values of $|l'|$, the elements of the off-diagonal blocks of $\mathbf{T}_{nn'}^{ll'}$ grow exponentially. 
On the one hand, for large values of $|l|$, the elements of the off-diagonal blocks of $\mathbf{T}_{nn'}^{ll'}$ \emph{decay} exponentially. 
The net result is that, for small values of $|l-l'|$, the elements of the off-diagonal blocks remain bounded, as the exponential growth with respect to $|l'|$ is compensated by the exponential decay with respect to $|l|$.
On the other hand, for large values of $|l-l'|$ \emph{and} large $|l'|$, the matrix elements quickly blow up.
This exponential growth can be clearly seen in Fig. \ref{fig:tmatrix}b for off-diagonal elements.
The consequences are important, since it is theoretically not possible to minimize the error of 2D-GLMT by solving arbitrarily large matrices because it would involve arbitrarily large numbers.

To avoid numerical instabilities, one can follow the rule of using ``not too large'' truncation orders. 
For a given cylinder array, the most common prescription for the truncation order $l_{\mathrm{max}}$ is \cite{Elsherbeni1992, Felbacq1994}
\begin{equation}
l_{\mathrm{max}} \sim 3 k r_{\mathrm{max}},
\end{equation}
where $r_{\mathrm{max}}$ is the radius of the largest cylinder in the array. 
With this rule, it is usually possible to solve scattering and resonances problems with an acceptable degree of accuracy.

Despite the fact that the prescription of using ``not too large'' truncation orders is adequate in many instances (see Ref. \cite{Elsherbeni1992, Felbacq1994, Natarov2014} and references therein), a robust method guaranteeing the convergence of the matrix equations to an \emph{arbitrary} truncation order would be more satisfactory. The recent formulation of Natarov \emph{et al.} achieves just that.
It consists in recasting the 2D-GLMT equation \eqref{eq:system} in a block-type Fredholm form of the second-kind.
One re-normalizes the beam shape coefficients $a_{nl}^0$ and the unknowns $b_{nl}$ using the formulas \cite{Natarov2014}
\begin{equation}
\begin{aligned}
a_{nl}^0 &= \hat{a}_{nl}^0 J_l(k_0 r_n) \\
b_{nl} &= \hat{b}_{nl} J_l(k_0 r_n),
\end{aligned}
\end{equation}
and solves the linear system for the new unknowns $\hat{b}_{nl}$. Substituting in \eqref{eq:bnl} yields
\begin{widetext}
\begin{equation}\label{eq:bnl2}
\hat{b}_{nl} - {s_{nl}}\sum_{n' \neq n} \sum_{l'=-\infty}^{\infty}e^{i(l'-l) \phi_{nn'}} \h_{l-l'}(k_0R_{nn'})\frac{J_{l'} (k_0 r_{n'})}{J_l(k_0 r_n)} \hat{b}_{n'l'} = {s_{nl}} \hat{a}_{nl}^0.
\end{equation}
The modified matrix equation is therefore
\begin{equation}\label{eq:system2}
\mathbf{\hat{T}} \mathbf{\hat{b}} = \mathbf{\hat{a}}_0,
\end{equation}
with
\begin{equation}\label{eq:tmatrix2}
\mathbf{\hat{T}}_{nn'}^{ll'} = \delta_{nn'}\delta_{ll'} - (1 - \delta_{nn'}) e^{i(l'-l) \phi_{nn'}} \h_{l-l'}(k_0R_{nn'}) \frac{J_ {l'}(k_0 r_{n'})}{J_l(k_0 r_n)} s_{nl}.
\end{equation}
\end{widetext}
The net effect of this re-normalization is that, for large $|l'|$, the additional factor exponentially decays and compensates the exponential growth of the Hankel function with respect to that index. 
Conversely, for large $|l|$, the additional factor exponentially grows and compensates the exponential decay with respect to that index.
Following this rescaling, it can be shown that a truncated version of \eqref{eq:system2} converges to the exact solution of the scattering/eigenmode problem for $l_{\max} \rightarrow \infty$ \cite{Natarov2014}.
In mathematical terms this stems from the fact that
\begin{equation}
\sum_{l,l'=-\infty}^{\infty} \left| \mathbf{\hat{T}}_{nn'}^{ll'} - \delta_{nn'}\delta_{ll'} \right|^2 < \infty, \qquad \forall~(n,n').
\end{equation}
In other words, the norm of the transfer matrix is finite regardless of the chosen truncation order $l_\mathrm{max}$.
Indeed, as can be seen from Fig. \ref{fig:tmatrix}c, the largest elements of off-diagonal blocks, transformed to a Fredholm form, are located near the diagonal of the blocks, whereas matrix elements remain bounded for large values of $|l - l'|$.
This was not the case for the original system \eqref{eq:system}, as evidenced by Fig. \ref{fig:tmatrix}b.

For simplicity, we shall use the non-renormalized form of the central equation \eqref{eq:system} of 2D-GLMT, unless indicated otherwise.
Clearly, the theoretical discussion holds for the normalized version \eqref{eq:system2} as well.
The computations can always be carried out for $\mathbf{\hat{T}}$ and $\mathbf{\hat{b}}$ instead of  $\mathbf{T}$ and $\mathbf{b}$.
Afterwards, one simply maps back the coefficients $\hat{b}_{nl} \rightarrow b_{nl}$ to reconstruct the electromagnetic fields.

\subsection{Scattering of arbitrary beams}
\label{sec:scattering}
The interaction of an incident beam -- in principle of arbitrary shape -- with an array of cylindrical scatterers can be readily modelled using 2D-GLMT.
Given a set of beam-shape coefficients $\lbrace a_{nl}^0 \rbrace$, the scattered coefficients $\lbrace b_{nl} \rbrace$ can be directly computed by solving the system of linear equations \eqref{eq:system}. 
In fact, as stated by Gouesbet and Lock, the main difficulty behind Lorenz-Mie theories is usually the computation of the beam-shape coefficients for a given excitation \cite{Gouesbet2013}.
Although the calculation of these coefficients is not our primary topic, it is an important ingredient for a complete implementation of the method. We therefore dedicate the next two sections to the derivation of the beam-shape coefficients for two typical cases, namely a plane wave (section \ref{sec:plane}) and a focused beam (section \ref{sec:csb}) parametrized by the complex-source beam (CSB) technique.

\subsubsection{Beam-shape coefficients for a plane wave}
\label{sec:plane}
Consider the following incident field (plane wave) on the scatterer array
\begin{equation}\label{eq:pw}
\varphi_i (\mathbf{r}) = \varphi_0 e^{i \mathbf{k} \cdot \mathbf{r}}.
\end{equation} 
The incident wavevector $\mathbf{k}$ is defined by a modulus $k_0 = k \sqrt{\epsilon_0}$ and an angle of incidence $\Theta$. 
The position vector $\mathbf{r}$ is defined with respect to the origin of a global coordinate system. 
In the frame of reference of the $n^\mathrm{th}$ scatterer, one writes
\begin{equation}\label{eq:pw2}
\varphi_i (\mathbf{r}) = \varphi_0 e^{i \mathbf{k} \cdot (\mathbf{R}_n + \mathbf{r}_n)}
= \varphi_0 e^{i \mathbf{k} \cdot \mathbf{R}_n} e^{ik_0 \rho_n [ \cos ( \theta_n  - \Theta) ]},
\end{equation} 
and with the Jacobi-Anger expansion \cite[p.~361]{Abramowitz1970}, one further has
\begin{equation}\label{eq:angers}
e^{ik_0 \rho_n [ \cos ( \theta_n  - \Theta) ]} = \sum_{l=-\infty}^{\infty} i^l J_l (k_0 \rho_n) e^{il (\theta_n - \Theta)}.
\end{equation}
The substitution of \eqref{eq:angers} in \eqref{eq:pw2} and comparison with \eqref{eq:field} yields the beam-shape coefficients, identical to those found in \cite{Elsherbeni1992, Nojima2005, Wait1955}
\begin{equation}
\label{eq:anl0pw}
a^0_{nl} = \varphi_0 e^{i \mathbf{k} \cdot \mathbf{R}_n} i^l e^{-il\Theta}.
\end{equation}

\subsubsection{Beam-shape coefficients for a complex-source beam}
\label{sec:csb}
The basic form of a two-dimensional Gaussian beam (GB) propagating along the $x$ axis, which satisfies the paraxial 2D Helmholtz equation, is given by \index{Beam!Gaussian}
\begin{equation}\label{eq:GB}
\varphi_g(x,y) = \sqrt{\frac{2}{\pi k_0 (x-x_\mathrm{R})}} \exp \left\lbrace ik_0 \left( x + \frac{1}{2} \frac{y^2}{x-ix_\mathrm{R}} \right)\right\rbrace.
\end{equation}
The parameter $x_\mathrm{R}$ is the Rayleigh distance of the GB and is proportional to the coherence length of the beam. The beam waist is located in the $x=0$ plane. The physical appeal and usefulness of the Gaussian beam is well established.
It is compatible with semi-analytical approaches such as ray-transfer (or $ABCD$) matrices and leads to a closed form solution of the Fresnel-Kirchhoff integral \cite{Saleh2007}.
A Gaussian beam is also a good approximation of the radiation pattern of the fundamental mode of a rectangular waveguide  \cite[pp.~43--46]{Okamoto2006} or of an optical fiber \cite{Marcuse1978}. 
However, the Gaussian beam solution suffers from two main drawbacks, a conceptual and a technical one.
First, it is but an approximate solution of the Helmholtz equation.
One must use the \emph{paraxial approximation} to obtain \eqref{eq:GB} from \eqref{eq:helmholtz}.
Second, computing the beam-shape coefficients for the canonical Gaussian beam solution is not easily accomplished.

To circumvent these two obstacles, one can use the complex-source beam (CSB) solution of the 2D Helmholtz equation as a closed-form incident field.
This solution has been proposed in order to extend the validity of the GB beyond the paraxial zone and exhibits the required cylindrical symmetry.
Using the Green's function of the inhomogeneous Helmholtz equation for a point source located in complex space at coordinates $x'=ix_\mathrm{R}$ and $y'=0$, one obtains the CSB solution \cite{Mahillo-Isla2008}
\begin{equation}\label{eq:CSB}
\varphi_i(\mathbf{r}) = \h_0 (k_0 r_s),
\end{equation} 
with
\begin{equation}
r_s \equiv [(y-y')^2 + (x - x')^2]^{1/2}= [y^2 + (x - ix_\mathrm{R})^2]^{1/2}.
\end{equation}
This solution is continuous everywhere in the real plane except across the branch cut connecting the two singularities at $(x,y) = (0,x_\mathrm{R})$ and $(x,y) = (0,-x_\mathrm{R})$.
The waist plane is thus located in the branch cut.
It should be noted that to obtain a regular solution in that plane, one can combine linearly independent CSB solutions as described in \cite{Mahillo-Isla2008}.
\begin{figure*}
	\centering
	\includegraphics[width=\textwidth]{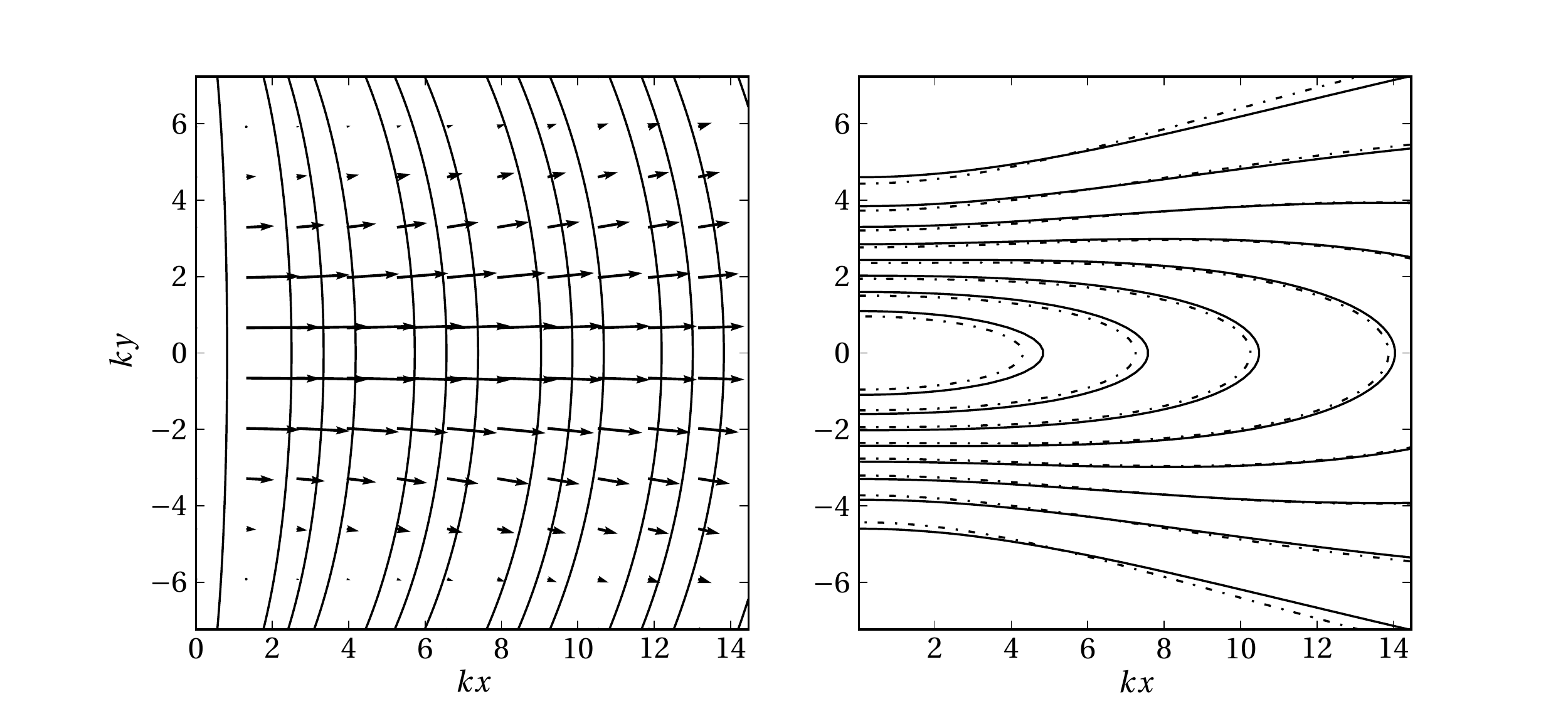}
	\caption[Phase fronts and Poynting vectors of the free-space CSB solution described by Eq. (\ref{eq:CSB})]{(Left) Phase fronts and Poynting vectors of the free-space CSB solution described by Eq. (\ref{eq:CSB}). (Right) Amplitude contours of the CSB solution and comparison with the canonical Gaussian beam (dashed). The normalized Rayleigh distance is set to $k_0 x_\mathrm{R}=9.65$, the same value used in \cite{Gagnon2012}. Reproduced from \cite{Gagnon2014th}.}\label{fig:csb}
\end{figure*}

The CSB solution reduces, up to a multiplicative constant, to a canonical Gaussian beam in the paraxial zone $x \gg y$. To show this, one can rewrite $r_s$ in the following way
\begin{equation}
r_s = \pm (x - ix_\mathrm{R}) \sqrt{1 + \dfrac{y^2}{(x - ix_\mathrm{R})^2}}, \qquad x \gtrless 0,
\end{equation}
and use the binomial approximation \cite{Heyman2001} to write
\begin{equation}
r_s \sim \pm \left[ (x - ix_\mathrm{R}) + \dfrac{1}{2} \dfrac{y^2}{x- ix_\mathrm{R}} \right], \qquad x \gtrless 0.
\end{equation}
Moreover, using the asymptotic expansion of Hankel functions for large arguments, one can write \cite{Abramowitz1970}
\begin{widetext}
\begin{subequations}\label{eq:gaussian}
\begin{equation}
\begin{aligned}
\varphi_i(x,y) 
& \sim \sqrt{\dfrac{2}{\pi k_0 r_s}} \exp{i \left( k_0 r_s  + \frac{\pi}{4} \right)  }  &\\
& \sim \sqrt{\frac{2}{\pi k_0 (x-ix_\mathrm{R})}} \exp ik_0 \left( x + \frac{1}{2} \frac{y^2}{x-ix_\mathrm{R}} \right)
e^\frac{i\pi}{4} e^{k_0 x_\mathrm{R}}, & \qquad x > 0,
\end{aligned}
\end{equation}
\begin{equation}
\begin{aligned}
\varphi_i(x,y) & \sim \sqrt{\frac{2}{\pi k_0 (x-ix_\mathrm{R})}} \exp -ik_0 \left( x + \frac{1}{2} \frac{y^2}{x-ix_\mathrm{R}} \right)
e^\frac{i\pi}{4} e^{-k_0 x_\mathrm{R}}, & \qquad x < 0.
\end{aligned}
\end{equation}
\end{subequations}
\end{widetext}
Eq. \eqref{eq:gaussian} and Fig. \ref{fig:csb} clearly show the transverse Gaussian profile of the CSB in the paraxial zone. 

The next step in the derivation of the beam shape coefficients for a focused beam is to expand the CSB on a basis of cylindrical waves centred on each individual scatterer. 
One rewrites \eqref{eq:CSB} as
\begin{equation}
\varphi_i(\rho_n,\theta_n) =  \h_0 (k_0 |\mathbf{r}_n - \mathbf{r}_{sn}|),
\end{equation}
where $\mathbf{r}_n = (\rho_n,\theta_n)$ and $\mathbf{r}_{sn}$ is the vector pointing from the center of the $n^\mathrm{th}$ scatterer to the complex source point. We apply Graf's addition theorem with the following change of variables
\begin{equation}
U = k_0 r_{sn}, V = k_0 \rho_n, W = k_0 |\mathbf{r}_n - \mathbf{r}_{sn}|,
\end{equation} 
where $\rho_n = |\mathbf{r}_n|$ and $r_{sn} = |\mathbf{r}_{sn}|$ (Euclidean norms). In accordance with \cite{Oguzer1995}, we write
\begin{subequations}
\begin{equation}
 - \mathbf{r}_{sn} = (X_n - ix_\mathrm{R}) \hat{\mathbf{e}}_x + Y_n \hat{\mathbf{e}}_y,
\end{equation}
\begin{equation}\label{eq:rsn}
 r_{sn} = \sqrt{(X_n - ix_\mathrm{R})^2 + Y_n^2},
\end{equation}
\end{subequations}
thus
\begin{subequations}
 \begin{equation}
\mathbf{r}_n - \mathbf{r}_{sn} = (x_n + X_n - ix_\mathrm{R})\hat{\mathbf{e}}_x + (y_n + Y_n) \hat{\mathbf{e}}_y,
\end{equation}
\begin{equation}
 |\mathbf{r}_n - \mathbf{r}_{sn}| = \sqrt{(x_n + X_n - ix_\mathrm{R})^2 + (y_n + Y_n) },
\end{equation}
\end{subequations}
where $(x_n,y_n)$ are Cartesian coordinates centred on the $n^\mathrm{th}$ scatterer. Using \eqref{eq:grafcv}, one obtains
\begin{equation}
 \cos(\alpha_n + \pi) = \cos(\theta_n - \mu_n),
\end{equation}
where 
\begin{subequations}
 \begin{align} \label{eq:mu}
 \cos \mu_n = \frac{X_n - ix_\mathrm{R}}{r_{sn}} \qquad &, \qquad  \sin \mu_n = \frac{Y_n}{r_{sn}}, \\
 \cos \theta_n = \frac{x_n}{\rho_n}   \qquad &, \qquad  \sin \theta_n = \frac{y_n}{\rho_n}.
\end{align}
\end{subequations}
Using these substitutions, the expansion takes the form
\begin{equation}\label{eq:simple}
  \varphi_i(\rho_n,\theta_n) = \sum_{l=-\infty}^{\infty} \h_l(k_0 r_{sn}) J_l(k_0 \rho_n) e^{i l \alpha_n},
\end{equation}
and substituting $\alpha_n = \theta_n - \mu_n - \pi$, one finally obtains the beam-shape coefficients for a CSB
\begin{equation}\label{eq:oguzer}
 a_{nl}^0 = (-1)^l \h_l(k_0 r_{sn}) e^{-il\mu_n},
\end{equation}
where $r_{sn}$ is given by \eqref{eq:rsn} and $\mu_n$ by \eqref{eq:mu}.

The expansion (\ref{eq:simple}) is similar to Eq. (3) in reference \cite{Oguzer1995}, with the difference that the condition $|U e^{\pm i \alpha_n}| < |V|$ is required in \cite{Oguzer1995} because the source is located inside a circular reflector, whereas \eqref{eq:oguzer} requires $|V e^{\pm i \alpha_n}| < |U|$. 
In accordance with this condition, the convergence of (\ref{eq:simple}) is limited to a disk not intersecting or touching the branch cut between $(x,y)=(0,x_\mathrm{R})$ and $(x,y)=(0,-x_\mathrm{R})$. In other words, scatterers must not intersect or touch the branch cut for the expansion to hold.
This condition is easily satisfied in any given setting.

\subsection{Eigenmode computations}
\label{sec:eigenmodes}

In the previous section, we have shown that the scattering wavefunctions of an array of cylinders can be computed by solving a linear system of equations. 
Eigenmode computations involve instead a homogeneous version of this linear system. 
Generally, this system is of the form
\begin{equation}\label{eq:lasing}
\mathbf{T}(\lambda) \mathbf{b} = 0,
\end{equation}
i.e. the computation of lasing states is a non-linear eigenvalue problem.
One must compute the discrete set of (generally complex) eigenvalues $\lambda$ and eigenvectors $\mathbf{b}$ for which $\det(\mathbf{T}) = 0$.
Since the non-linear eigenvalue is usually a complex frequency or wavenumber, this problem amounts to finding the resonant frequencies associated to an \emph{infinite} scattered amplitude in the presence of a \emph{finite} amplitude incident wave.
Algorithms for solving the non-linear eigenvalue problem are described in \cite{Heider2010, Ruhe1973, Wiersig2003}.

We now show that \eqref{eq:lasing} can be readily adapted for the computation of the classical QB states of an array of cylinders, and of the CF states described in section \ref{sec:cf}.
The QB states of an array of dielectric scatterers satisfy the following Helmholtz equation
\begin{subequations}
\begin{align}
[\nabla^2 + \epsilon_0 k^2]\varphi(\rho_n,\theta_n) &= 0, \qquad (\mbox{Outside all cylinders}) \\
[\nabla^2 + \epsilon_n k^2]\varphi(\rho_n,\theta_n) &= 0, \qquad \rho_n < r_n.
\end{align}
\end{subequations}
Consequently, the matrix equation describing the quasi-bound modes is simply
\begin{equation}\label{eq:qbmodes}
\mathbf{T}(k) \mathbf{b} = 0,
\end{equation}
with the substitution $k_n \leftarrow k \sqrt{\epsilon_n}$ in \eqref{eq:snl}.

As stated in section \ref{sec:salt}, QB states cannot accurately describe the steady-state lasing behavior of an array of active cylinders, even near threshold \cite{Ge2010}. This is due to the QB eigen-frequencies being complex everywhere outside the cylinders, resulting in exponential growth of the electromagnetic energy at infinity \cite{Ge2010}. To enable a more realistic treatment of eigenmodes, we have introduced in section \ref{sec:cf} a new kind of eigenstate central to SALT, the CF state. One of the appealing features of CF states is that they are readily computed using the Lorenz-Mie approach, under the single assumption that the cavity region is composed of a subset of the cylinder array. In other words, the medium surrounding all cylinders is passive, and some cylinders may also be passive. \index{Medium!passive} Accordingly, the wavevector is complex only inside active cylinders. The CF states therefore satisfy
\begin{subequations}
\begin{align}
[\nabla^2 + \epsilon_0 k^2]\varphi &= 0, \qquad (\mbox{Outside all cylinders}) \\
[\nabla^2 + \epsilon_n k^2]\varphi &= 0, \qquad (\mbox{Inside passive cylinders}) \\
[\nabla^2 + \epsilon_n K(k)^2]\varphi &= 0, \qquad (\mbox{Inside the cavity region}).
\end{align}
\end{subequations}
The matrix equation describing the CF states is \index{Cavity!region}
\begin{equation}\label{eq:cfmodes}
\mathbf{T}(K) \mathbf{b} = 0,
\end{equation}
with the substitution $k_n \leftarrow K \sqrt{\epsilon_n}$ in \eqref{eq:snl} if the $n^\mathrm{th}$ cylinder is active, and  $k_n \leftarrow k \sqrt{\epsilon_n}$ otherwise. Note that, as always, the complex eigenfrequency \index{Eigenfrequency} $K$ associated to the CF states depends on the value of the exterior frequency \index{Frequency!exterior} $k$.

In summary, 2D-GLMT can be used to compute the eigenstates of arrays of coupled active cylinders, possibly all different in size and refractive indices. 
The method works equally well for the computation of the eigenfunctions and eigenfrequencies of QB states, meta-stable solutions of the 2D Helmholtz equation, and for the computation of CF states. 
These ``improved'' eigenstates central to the SALT theory are more physically realistic solutions to the Helmholtz equation in the sense that they remain bounded at infinity, unlike QB states. 
Using CF states also enables the computation of resonances of photonic complexes in the case where some cylinders are active and other remain passive. 
This is incompatible with QB states computations, which amount to an active medium extending to the whole spatial domain.

\section{Numerical examples}
\label{sec:examples}

This section is dedicated to numerical examples of both scattering and eigenmode computations using 2D-GLMT.
It serves as a synthesis of all the theoretical concepts discussed in the previous two sections, including scattering of focused beams by coupled cylinders and eigenstates drawn from the SALT formulations.

Two-dimensional photonic crystals (PhCs) based on arrangements of cylindrical scatterers are one of the prime candidates for modelling using 2D-GLMT \cite{Nojima2005, Gagnon2012, Gagnon2013, Gagnon2014b, Vukovic2010}.
Although PhCs are strictly speaking infinite periodic arrangements of scatterers, 2D-GLMT does not require a symmetric arrangement, nor is it meant to be applied to infinite structures.
With these caveats in mind, we will sometimes refer to finite, periodic or non-periodic configurations as ``PhC devices''.
In section \ref{sec:filter}, we present the numerical example of an integrated polarization filter as a representative scattering computation using 2D-GLMT.
In section \ref{sec:phc}, the focus is on the more subtle eigenmode computations.
These computations are presented through the example of a so-called ``photonic bandgap defect cavity''.

Although our examples are restricted to devices comprising a number of cylinders or cylindrical inclusions ranging from $\sim 50$ to several hundreds, they are by no means the only possible application of 2D-GLMT.
Other notable instances of photonic and plasmonic devices that have been successfully modeled using 2D-GLMT include photonic molecules \cite{Smotrova2006, Smotrova2013, Gagnon2014a}, integrated silicon lenses \cite{Sanchis2004, Marques-Hueso2013}, nanowires \cite{Natarov2014} and coupled-resonator optical waveguides \cite{Pishko2007}. 
2D-GLMT can also be extended to geometries characterized by an infinite extent in the $z$ direction, such as micro-structured optical fibres (MOFs).
The analytic formulation for MOFs can be found in \cite{White2002} with details on the implementation and numerical examples in \cite{Kuhlmey2002}.
In this ``extended'' formulation, the complex eigenfrequency $k_\mathrm{QB}$ of QB states described by equation \eqref{eq:helmholtz} can be interpreted as a ``transverse'' propagation constant which can be used to compute a ``longitudinal'' propagation constant $\beta$ using $\beta^2 = k_0^2 - k_\mathrm{QB}^2$, where $k_0$ is the free-space wavevector.
The modelling of MOFs constitute an important research topic, since these fibres can be used to accelerate charged particles using TM-polarized photonic bandgap defect modes analogous to those computed in section \ref{sec:phc}.
TM modes in hollow-core fibres are particularly appealing for electron acceleration since they exhibit only a longitudinal electric field parallel to the fibre axis \cite{Noble2011}.

\subsection{Scattering computations: polarization filters}
\label{sec:filter}
Our numerical application of a scattering computation is the characterization of integrated polarization filters based on engineered photonic lattices.
This example serves two additional purposes.
It illustrates the possibility to use focused beams described by Eq. \eqref{eq:CSB} in scattering computations.
It also emphasizes the fact that the numerical approach described in section \ref{sec:glmt} allows to treat both orthogonal polarization components (TE and TM) independently.

Our polarization filter is composed of cylindrical inclusions of refractive index $n=1$ embedded in a medium of refractive index $n=2.76$, corresponding to the effective value for a thin silicon slab.
The radius of the inclusions is set to $r = 0.3 \Lambda$, where $\Lambda$ is the lattice constant.
This type of PhC geometry can be generically referred to as holes-in-slab (HIS).
Although the filter is finite and non-periodic, its infinite counterpart -- a square HIS lattice -- exhibits a directional bandgap for the TE polarization around the normalized frequency $k_0 \Lambda = 1.5$.
This implies that the TE component of the beam ($E_y$) is more strongly scattered, suggesting HIS based designs are suitable for filtering this polarization out and favouring the TM polarization.
This bandgap analysis, based on the plane wave expansion method, is detailed in \cite{Gagnon2014b}.

\begin{figure}
	\centering
	\includegraphics{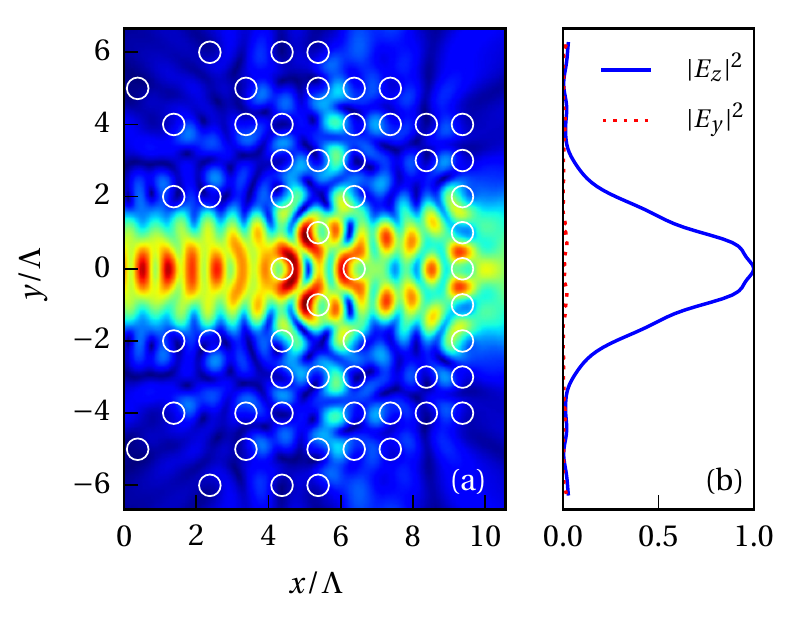}
	\caption{Generation of a TM polarized Gaussian beam. (a) Optimized configuration (57 scatterers) and $|E_z|$ field profile (arbitrary units). The target plane coincides with the upper limit of the $x$ axis. (b) Comparison of orthogonal polarization components along target plane. The energy of the polarized component of the beam is $\sim 60$ times higher than that of the unpolarized component. Configuration detailed in Ref. \cite{Gagnon2014b}.}\label{fig:polarization}
\end{figure}

Using optimization methods presented in \cite{Gagnon2013, Gagnon2014b}, one is able to engineer a lattice configuration for simultaneous polarization filtering and beam shaping, that is obtaining a given beam shape after filtering.
A configuration optimized for the conversion of an incident Gaussian beam to a identical -- but polarized -- Gaussian beam is shown in Fig. \ref{fig:polarization}.
The configuration of cylindrical holes is shown on Fig \ref{fig:polarization}a, along with a false colour plot of the field amplitude inside the device.
In this example, the incident beam wavenumber is set to $k_0 \Lambda = 1.76 $, chosen to fall in the vicinity of the directional bandgap of the infinite HIS lattice.
Both the output and input beams' half-width are set to $w_0=2.5 \Lambda$.
The obtained output beam, shown in Fig. \ref{fig:polarization}b, deviates from a Gaussian amplitude profile by about $4 \%$ (RMS value), while the energy of the polarized component of the beam ($E_z$) is $\sim 60$ times higher than that of the non-polarized component ($E_y$).
This corresponds to a degree of polarization of more than 98 \%.
More details about the configurations, as well as the optimization of a TE polarizer, can be found in \cite{Gagnon2014b}.

A few remarks concerning the scattering computation shown in Fig. \ref{fig:polarization} are in order.
We did not use the renormalized version of the matrix for scattering computations since it was not required for a numerically stable solution.
Moreover, as stated in section \ref{sec:csb}, the complex-source beam approach precludes the positioning of scatterers in the beam waist.
Therefore, we positioned all scatterers at $x > r$.
For well-collimated beams, this is not a critical impediment.

To illustrate the full potential of 2D-GLMT for photonic structures characterization and optimization, it is instructive to consider the numerical cost of a computation such as the one shown in Fig. \ref{fig:polarization}.
The dimension of the underlying transfer matrix is $\sigma = N (2 l_\mathrm{max} + 1)$, where $l_\mathrm{max} = 3 \mathrm{int} (k r) + 1 = 2$.
Since this polarization filter is composed of $N=57$ cylindrical scatterers, this translates to $\sigma = 285$.
In other words, the computation of the scattered field shown in Fig. \ref{fig:polarization} implies the solution of a square linear system of 285 equations, or equivalently the inversion of a $285 \times 285$ complex matrix.
On a personal computer, this operations takes a mere fraction of a second.
Storing the larger transfer matrices associated to high $N k r$ values is also accessible, but may benefit from the use of parallel codes.
Although this may seem restrictive, we stress the fact that large $Nkr$ values do not lead to stability issues, especially considering the improvement presented in section \ref{sec:fredholm} which ensures that transfer matrices remain well-conditioned regardless of their size.

The speed of the 2D-GLMT method is critical if the goal of a research project is the \emph{optimization} of a photonic device.
Indeed, the fast characterization of cylinder arrays enabled by 2D-GLMT allows it to be coupled with \emph{metaheuristic algorithms}, such as the genetic algorithm (GA).
The combination of 2D-GLMT + GA has been successfully used in the past for the design of waveguide bends \cite{Vukovic2010}, beam shapers \cite{Gagnon2012} and integrated lenses \cite{Marques-Hueso2013} based on cylindrical scatterers.
More generally, 2D-GLMT can be coupled with any optimization algorithm able to deal with a binary encoding of the problem.
The scatterer configuration of Fig. \ref{fig:polarization}a was actually found using a combination of 2D-GLMT and an algorithm called \emph{tabu search}.
More details can be found in two of our recent publications \cite{Gagnon2013, Gagnon2014b}.
Continuous optimization algorithms such as inverse harmony search \cite{Andoneguii2013} may also be combined with 2D-GLMT.
For instance, one might decide to optimize not only the presence/absence of a scatterer, but also its geometric parameters such as its radius and refractive index.

\subsection{Eigenmode computations: photonic crystal cavities}
\label{sec:phc}
We now present numerical examples of CF states of coupled cylinder arrays
\footnote{For all eigenmode computations presented in this tutorial, we use the renormalized version of the 2D-GLMT equations, as described in section \ref{sec:fredholm}. This ensures a higher precision on the resonant frequencies and field patterns.}.
Specifically, computing CF states using 2D-GLMT implies searching for eigenfrequencies $K_\mu$ solutions of \eqref{eq:cfmodes} and the associated eigenfunctions $\varphi_\mu$ using the scheme described in Appendix \ref{sec:computation}.
A countably infinite set of eigenvalue-eigenfunction pairs can be computed for every different value of the real exterior frequency $k$, but only discrete values of $k$ will yield a physically realistic TLM, as described in section \ref{sec:salt}.
We shall not map CF states on individual TLMs, since in the case of uniform pumping this mapping is straightforward.
We refer interested readers to one of our previous articles for more details on this mapping \cite{Gagnon2014a}.

\begin{figure}
	\centering
	\includegraphics[width=3in]{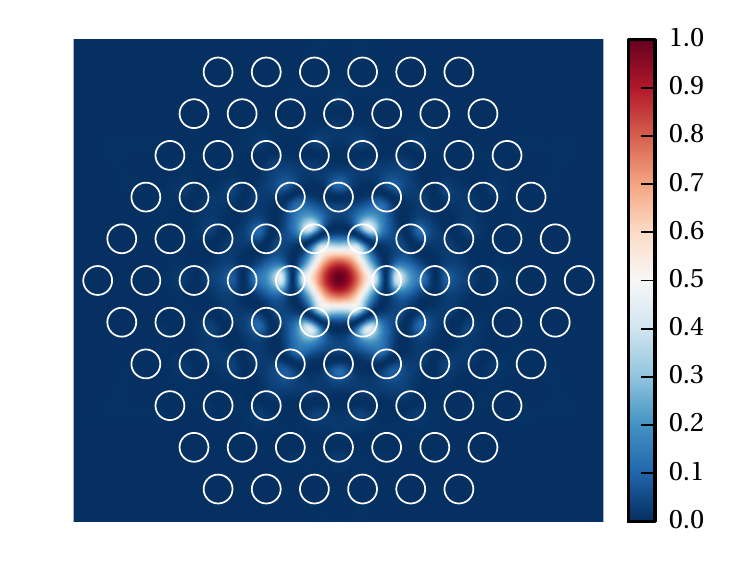}
	\includegraphics[width=2.97in]{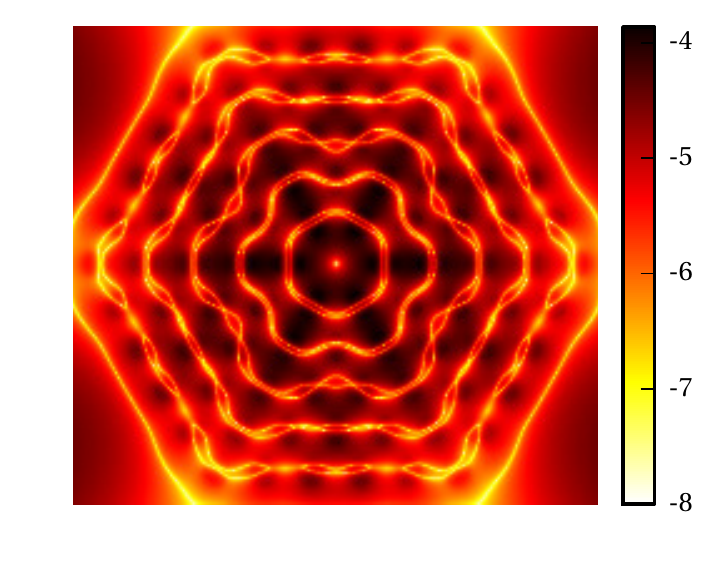}
	\caption{(Top) Amplitude profile $|\varphi|^2$ of a CF state of a photonic crystal defect cavity for an exterior frequency $k\Lambda=1.885$. The eigenvalue of this mode is $K\Lambda=1.885 -0.0044i$. (Bottom) Absolute value of the difference between the profile of the CF state and a neighbouring QB state with eigenfrequency $k\Lambda=1.885 -0.0035i$, or $Q \sim 260$ (log scale).} \label{fig:hex}
\end{figure}

A simple photonic crystal cavity is used as a representative cylinder geometry.
Unlike a typical Fabry-P\'erot where confinement is achieved via total internal reflection, \emph{bandgaps} are responsible for confinement in a photonic crystal laser.
The bandgaps are forbidden frequency ranges arising from scattering and interference in a periodic structure.
Introducing a \emph{defect} in such a periodic structure may result in photons remaining confined inside, creating a so-called ``photonic bandgap defect mode'', which is non-propagating.
We focus our attention on point defects, as experimentally reported by Painter \emph{et al.} \cite{Painter1999}.

The cavity geometry used is composed of 90 identical rods with a passive dielectric permittivity $\epsilon_c = 13.18$.
The rods are arranged on a triangular lattice of which we consider only a hexagonal subset with a defect at its centre (see the overlay on Fig. \ref{fig:hex}).
The radius of the rods is set to $r=0.3 \Lambda$, where $\Lambda$ is the lattice spacing.
The infinite counterpart of this finite photonic crystal array exhibits a complete bandgap for the TM polarization \cite{Aryal2009}.

Before searching for the CF states of this cylinder array, it is useful to compute the QB states first since they can be associated one-to-one with CF states.
It is subsequently straightforward to look for CF states using the fact that $K_\mu(k')$ is usually located close to $k_{\mathrm{QB}}$ in the complex plane \cite{Ge2010th}.
A false colour plot of a CF state found using this prescription is shown in Fig. \ref{fig:hex}.
One can see that the energy of the eigenmode is localized close to the point defect.
The eigenvalue of this CF state ($K\Lambda=1.885 -0.0044i$) falls into the complete bandgap of the triangular rod lattice described earlier.
On the bottom panel of Fig. \ref{fig:hex}, we display the difference between the amplitude profile of this CF state and the associated QB state; it is smaller than $10^{-4}$ for the range considered.
Interestingly, one can see that the CF and QB states tend to differ mostly outside the contour containing all scatterers.
This is a testament to the fact that CF states remain bounded at infinity, unlike QB states, as also illustrated in Fig \ref{fig:qbcf}.
In short, this numerical example shows that 2D-GLMT is versatile enough for the computation of QB and CF eigenstates of arrays of cylinders.

\subsection{Random lasers}
\label{sec:random}
Random lasers \cite{Wiersma2008,Cao2009} are non-conventional resonant structures where the usual laser cavity is replaced by a strongly scattering arrangement of particles embedded in an active medium, or of active particles in a host medium.  
These novel lasers sources are appealing structures because of their potentially very small size, low-cost and ease of fabrication. 
The realization of random lasers using optically pumped active ``powders'' has been demonstrated experimentally \cite{Wiersma2008, Cao2009}.
It is also possible to fabricate disordered lasers in a more controlled way using photonic-crystal-like structures. This was recently achieved via the etching of cylindrical inclusions in a GaAs planar waveguide, optically activated by layers of InAs quantum dots \cite{Riboli2014}.
Several applications of random lasers have already been proposed, ranging from tumour detection \cite{Wiersma2008} to optical trapping \cite{Riboli2014}.

The recent interest in random lasers has given rise to a number of modelling tools designed to harness the modal characteristics of strongly scattering random media.
SALT, for instance, provides an ideal framework for the study of complex lasing media, as a series of recent publications indicate \cite{Andreasen2011, Andreasen2014, Bachelard2012, Tuereci2008, Tuereci2009}.
Moreover, the potential of 2D-GLMT for the computation of modes of uniformly pumped random lasers was first proposed in \cite{Andreasen2011}.

\begin{figure}
	\centering
	\includegraphics[width=3in]{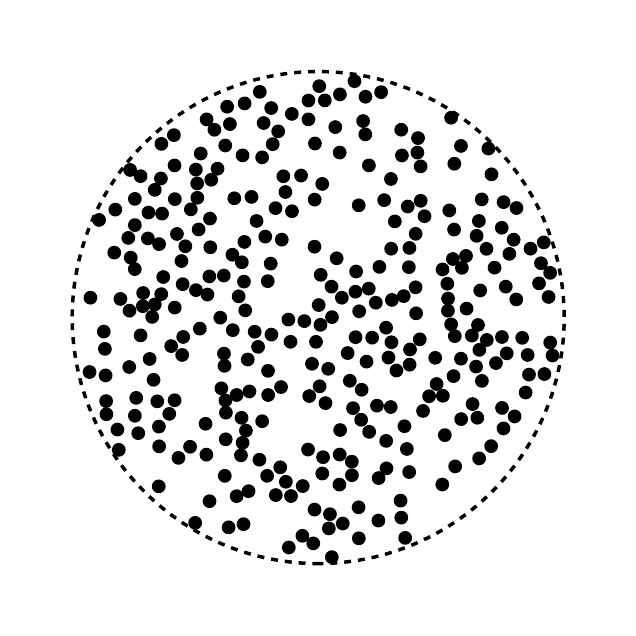}
	\caption{A possible realization of a random laser composed of 320 identical active cylinders with a surface filling factor of 0.2. Reproduced from \cite{Gagnon2014th}.} \label{fig:random}
\end{figure}

We now consider the computation of the CF states of a large array of active cylinders via 2D-GLMT.
We wish to illustrate the main numerical difficulties associated to the computation of CF and QB states of random lasers.
Using Ref. \cite{Fujii2012} as a guideline, we assume that an array of identical, non-overlapping rods are uniformly distributed within a circle of radius $R_{\mathrm{out}}$ with a predefined surface filling factor. For the purpose of illustration, we consider one realization of an array with $R_{\mathrm{out}} = 40$, normalized with respect to the radius of individual rods ($r=1$). We set the filling factor to 0.2, for an array containing 320 active scatterers with relative permittivity $\epsilon_c$ embedded in air ($\epsilon = 1$).
The geometry of the resulting array is shown in Fig. \ref{fig:random}. 
This arrangement of cylinders provides a useful, albeit simplified, model of a 2D random laser.

\begin{figure}
	\centering
	\includegraphics[]{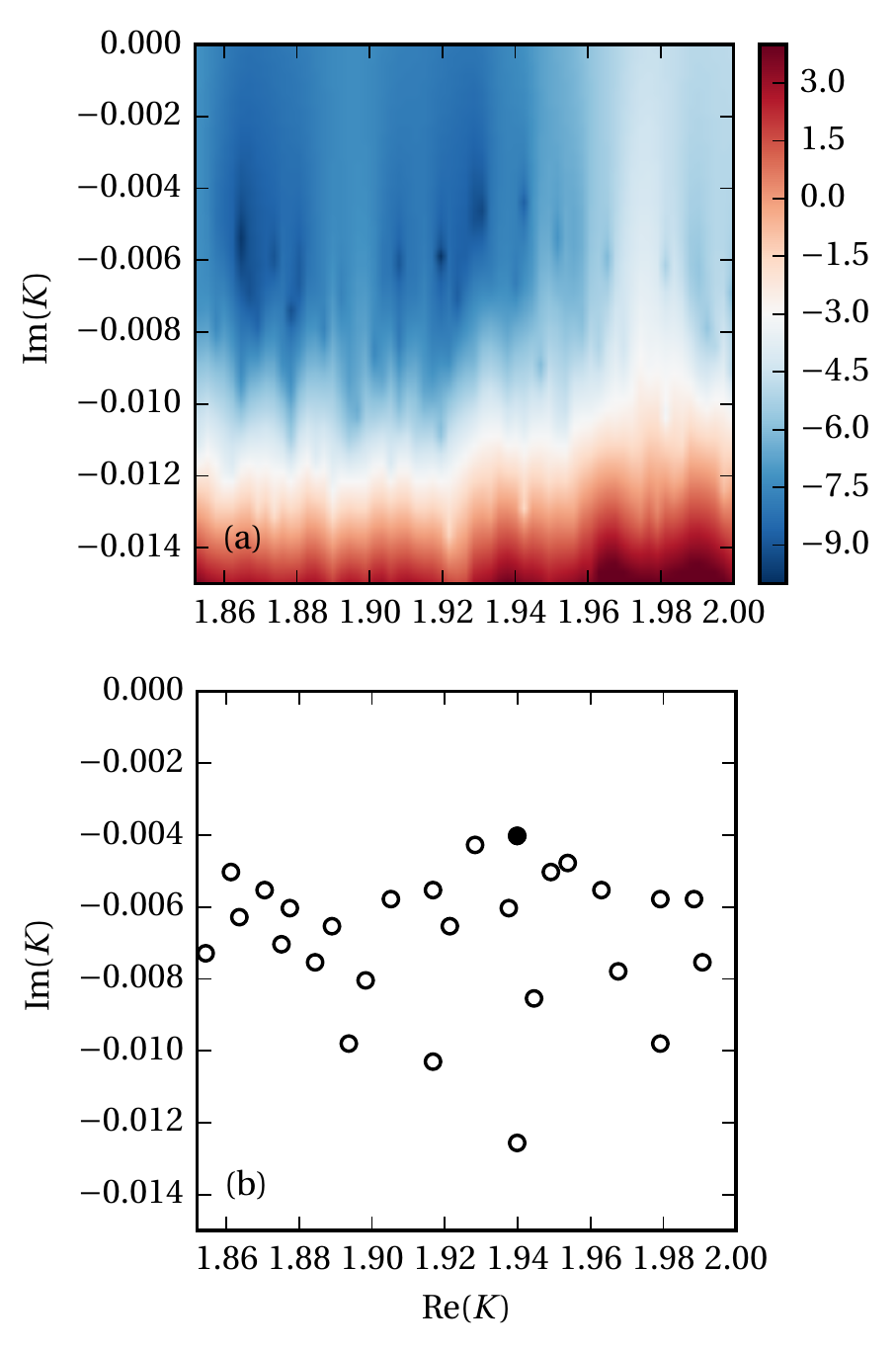}
	\caption{Determination of CF states for the geometry shown in Fig. \ref{fig:random}. (a) Evolution of $\log|\det[\mathbf{T}]|$ in the complex $K(k=2)$ plane. (b) Predictors of the position of various CF states of the random laser for $\epsilon_c = 4$. These correspond to local minima in the surface defined by $\log|\det[\mathbf{T}]|$. The solid dot indicates the position of the predictor of the CF state shown in Fig. \ref{fig:cfstate}.} \label{fig:detgrid}
\end{figure}

\begin{figure}
	\centering
	\includegraphics[width=3in]{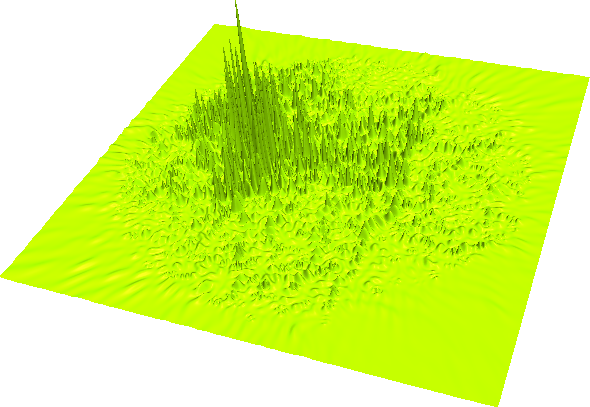} \\
	\caption{Field distribution $|\varphi|^2$ of a single CF state with $K = 1.9420 -0.004i$ $(Q \sim 240)$ for the geometry shown in Fig. \ref{fig:random} with $\epsilon_c = 4$. The height coordinate is proportional to the intensity (TM polarization).} \label{fig:cfstate}
\end{figure}

\begin{figure}
	\centering
	\includegraphics[width=3in]{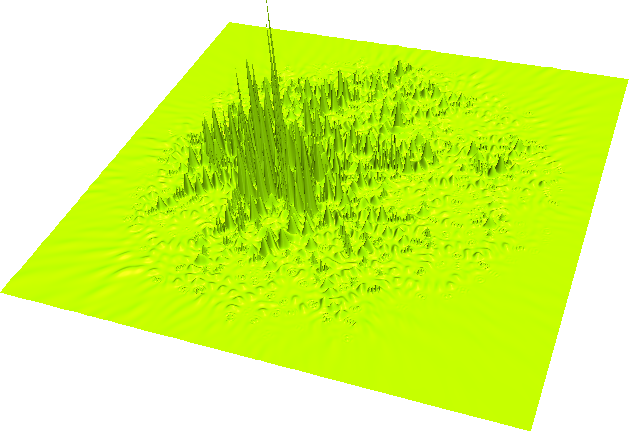} \\
	\caption{Field distribution $|\varphi|^2$ of a single CF state with $K = 1.9184 -0.002i$ $(Q \sim 480)$ for the geometry shown in Fig. \ref{fig:random} with $\epsilon_c = 9$. The height coordinate is proportional to the intensity (TM polarization).} \label{fig:cfstate2}
\end{figure}

The method for computing CF states described in Appendix \ref{sec:computation} can be easily applied to the random array without significant modification.
The main difficulty is rather numerical and lies in the much greater computation time for the determinant $\det[\mathbf{T}]$ of the transfer matrix for a large number of cylinders. 
Also, the density of states in a given wavenumber interval is greater for a large array.
Let us first set the permittivity of the cylinders to $\epsilon_c = 4$.
Using a sufficiently fine sampling of the complex $K$ plane (see Fig. \ref{fig:detgrid}a), we are able to obtain estimates for the position of a number of CF states (see Fig. \ref{fig:detgrid}b). 
These estimates can subsequently be refined using the procedure described in Appendix \ref{sec:computation}. 
The field distribution of one of these CF states is shown in Fig. \ref{fig:cfstate}.
For the parameters used (surface filling factor and refractive index), the scattering strength turns out to be relatively modest and the CF state intensity extends over a large fraction of the surface of the random laser. 
By increasing the scattering, for instance with a larger refractive index contrast between the cylinders and the surrounding medium and/or a bigger filling factor, the eigenmodes of the system can become localized in an effective closed cavity. 
This spatial confinement is the principal attribute of an {\em Anderson localized regime} whereby the light intensity occupies a smaller, ultimately microscopic, region of the random array \cite{Tuereci2008, Tuereci2009, Wiersma2013}.
As an illustration, consider the result shown in Fig. \ref{fig:cfstate2}, which consists in a CF state of the geometry shown in Fig. \ref{fig:random}, but for a value of $\epsilon_c = 9$.
The $Q$ factor of this resonance is approximately twice that of the CF state shown in Fig. \ref{fig:cfstate}, and the intensity outside the cavity is smaller with respect to the peak intensity of the mode ($Q = - \mathrm{Re}[K]/2\mathrm{Im}[K]$).
This is indicative of a transition towards a fully localized regime associated with minimal losses.
For a thorough discussion of Anderson localization of light, we refer the interested reader to a recent topical review on the subject \cite{Mafi2015}.

The computation of the field distribution with the resolution shown in Fig. \ref{fig:cfstate} and \ref{fig:cfstate2} is computationally intensive (several hours on a supercomputer) due to the large number of cylindrical harmonics that contribute to the total wavefunction at a given point in space ($2l_\mathrm{max} + 1$ for each scatterer).
The transfer matrices associated to this geometry imply also important memory requirements.
We should stress that eigenmode computations are always more numerically costly than scattering computations.
The reason is clear: eigenmode computations often imply some kind of wavenumber sampling (as shown in Fig. \ref{fig:detgrid}a), giving rise to a much larger number of complex matrix operations \cite{Kuhlmey2002}.

To conclude, a final point should be mentioned concerning a limitation of 2D-GLMT for the modelling of complex lasing media.
Throughout our presentation, we have only considered photonic media composed of \emph{homogeneous} cylinders, i. e. piecewise constant refractive index distributions.
However, inhomogeneous and partially pumped resonators have been the object of many recent articles \cite{Andreasen2014, Bachelard2012, Bachelard2014, Esterhazy2013, Hisch2013, Liertzer2012, Liew2014}.
The main conclusion of these publications, obtained using SALT, is that non-uniformly pumping a laser (random or not) is an ideal method to tune its emission properties.
This partial pumping corresponds to a dielectric distribution $\epsilon(\mathbf{r})$ that varies inside a single scatterer or resonator, possibly following a smooth function.
Modelling inhomogeneous or partially pumped optical media represents an additional challenge since the spatial discretization of inhomogeneous regions is unavoidable.
This unfortunately precludes the use of 2D-GLMT as presented here.
Despite this limitation, 2D-GLMT remains highly useful owing to its numerical stability (especially considering the improvement proposed in section \ref{sec:fredholm}) and its relative speed in the absence of spatial meshing.

\section{Summary}
We have presented a detailed theoretical description of 2D-GLMT, a numerical method dedicated to the modelling of scattering and eigenmodes of coupled cylinder structures.
This method aims to exploit the internal symmetry of multi-cylinder structures to allow the expansion of the electromagnetic field in a basis of cylindrical functions centred on every individual cylindrical scatterer.
After a brief presentation of the basic differential equations describing wave propagation in 2D passive (section \ref{sec:em}) and active (section \ref{sec:salt}) geometries, a detailed description of the 2D-GLMT equations was presented (section \ref{sec:basic}), the key part of which resides in the application of Graf's addition theorem for cylindrical functions.
We have also detailed how, after the application of appropriate boundary conditions at the interface of individual cylinders (section \ref{sec:boundary}), the 2D-GLMT equations can be reduced to matrix form.
A recent improvement of the method consisting in reformulating the matrix equations into a Fredholm form of the second-kind form was also presented (section \ref{sec:fredholm}).
This improvement ensures that the transfer matrices can be truncated to an arbitrary order while still providing accurate numerical solutions.

After this theoretical description, we have performed several numerical implementations to illustrate the potential of 2D-GLMT for the modelling of complex photonic media, as well as the computational challenges associated to the use of the algorithm.
These examples fall in two broad categories, namely scattering (section \ref{sec:scattering}) and eigenmode (section \ref{sec:eigenmodes}) computations.
The characterization of PhC-inspired polarization filters was detailed as a representative example of a scattering computation (section \ref{sec:filter}).
This example also served to highlight the speed of 2D-GLMT, a useful feature when used in tandem with numerical optimization algorithms.
This speed stems from the fact that 2D-GLMT does not involve a spatial discretization, or meshing, of the physical space of the problem, unlike the more general FEM and FDTD algorithms.
We have also presented a celebrated application of 2D-GLMT, namely the computation of modes of PhC cavities (section \ref{sec:phc}).
Finally, the computation of modes of random lasers was presented to push the algorithm to its numerical limits (section \ref{sec:random}).

\section*{Acknowledgements}

The authors are thankful to J. Dumont and J.-L. D\'eziel for useful discussions and comments, and to A. I. Nosich for introducing us to the normalization procedure described in Section \ref{sec:fredholm}.
We acknowledge financial support from the Natural Sciences and Engineering Research Council of Canada (NSERC).
D. G. thanks S. MacLean for continuing support.
Computations were made on the supercomputers \emph{Colosse} and \emph{Mammouth}, managed by Calcul Qu\'ebec and Compute Canada.
The operation of these computers is funded by the Canada Foundation for Innovation (CFI), Minist\`ere de l'\'Economie, de l'Innovation et des Exportations du Qu\'ebec (MEIE), RMGA and the Fonds de recherche du Qu\'ebec - Nature et technologies (FRQ-NT).
We are also happy to mention the useful free software projects \cite{Armadillo,Tange2011}.

\appendix
\section{Computation of eigenmodes}
\label{sec:computation}
As stated in section \ref{sec:eigenmodes}, the computation of QB states is a non-linear eigenvalue problem (NLEP) of the form
\begin{equation}\label{eq:nle}
\mathbf{T}(k) \mathbf{b} = 0,
\end{equation}
where $\mathbf{T}$ is a square complex matrix.
A similar system for CF states is given by \eqref{eq:cfmodes}.
NLEPs consists in finding the eigenvalues $k$ and eigenvectors $\mathbf{b}$ satisfying \eqref{eq:nle}.
The theory of linear systems prescribes that non-trivial solutions to this equation exist if and only if
\begin{equation}\label{eq3:det}
\det[ \mathbf{T}(k) ] = 0.
\end{equation}
This class of problems is termed ``non-linear'' since $\mathbf{T}$ depends non-linearly on $k$. While solvers are readily available for linear eigenvalue problems, algorithms for solving the NLEP are more complicated. It is possible to solve the NLEP either using Newton's method, or using linearisation methods that reduce the problem to a sequence of linear ones, as described in \cite{Heider2010, Ruhe1973}.

In our work, we use Newton's method to find the eigenmodes, as described in \cite{Ruhe1973}.
The goal is to find roots of \eqref{eq3:det} numerically. One iteration of Newton's method can be written as
\begin{equation}
k_{n+1} = k_n - \frac{\det[\mathbf{T} (k_n)]}{\det[\mathbf{T} (k_n)]'},
\end{equation} 
where $k_0$ is an initial guess for the eigenvalue (complex wavenumber) and prime symbols indicate the derivative with respect to $k$. The derivative is computed using Jacobi's identity \cite{Magnus1999}
\begin{equation}
\det[ \mathbf{T} ]'= \det[ \mathbf{T} ] \mathrm{tr}\left[ \mathbf{T} ^{-1} \mathbf{T}' \right],
\end{equation}
where $\mathbf{T}'$ is the element-wise derivative of $\mathbf{T}$ with respect to $k$, and $\mathrm{tr}$ denotes the trace. This allows one to rewrite one step of Newton's method as \cite{Wiersig2003}
\begin{equation}
k_{n+1} = k_n - \frac{1}{\mathrm{tr}[\mathbf{T} ^{-1}(k_n)\mathbf{T} '(k_n)]}.
\end{equation}
Newton's method is known to converge quadratically to the roots of the determinant if one is sufficiently close to the said roots. To ensure convergence, the initial guesses for the eigenvalues $k_0$ must be chosen carefully.
In our studies, we obtain these initial guesses by evaluating the value of $\det[ \mathbf{T}(k) ]$ in a region of interest of the complex $k$ plane on a rectangular grid, as shown in Fig. \ref{fig:detgrid} for instance.
Points of this grid corresponding to a change in sign of the function are subsequently used as initial guesses for Newton's method, which in turn implies a numerical construction of $\mathbf{T}'$.
An alternative approach to computing $\det[ \mathbf{T}(k) ]$ is to use a singular value decomposition (SVD) \cite{Magnus1999}.
Is consists in tracking the smallest singular value of $\mathbf{T}$ on a rectangular grid and using the minima of this ``function'' as initial guesses.
This SVD approach can be useful when a geometrical symmetry of the cylinder array results in several degenerate non-linear eigenvalues, which can lead to a ``false minimum'' in the mapping of $\det[ \mathbf{T}(k) ]$, as described in Ref. \cite{Kuhlmey2002}.

Once the roots of \eqref{eq3:det} have been found using the Newton algorithm, one simply uses a SVD routine to obtain the null eigenvector $\mathbf{b}$, to reconstruct the electromagnetic field profile of the eigenmode using, for instance, Eq. \eqref{eq:scat}.
This method is general for QB and CF states and is not limited to 2D-GLMT computations.
In fact, it can also be used in conjunction with other methods based on transfer matrices that imply NLEPs, for example the boundary element method (BEM) for cavity resonances \cite{Heider2010, Wiersig2003}.

\bibliography{biblio}
\bibliographystyle{ieeetr}

\end{document}